\documentclass[
 reprint,
superscriptaddress,
 amsmath,amssymb,
 aps,
]{revtex4-1}
\usepackage{graphics}
\usepackage{dcolumn}
\usepackage{bm}
\usepackage{epstopdf}
\usepackage{epsfig}
\usepackage{color}

\begin{document}

\title{Stimulated quantum phase slips from weak electromagnetic radiations in  superconducting nanowires}

\author{Amir Jafari-Salim}
 \email{ajafaris@uwaterloo.ca}
  \affiliation{Department of Electrical and Computer Engineering,
 University of Waterloo, Waterloo, Ontario, Canada N2L 3G1}
 \affiliation{Institute for Quantum Computing, Waterloo, Ontario, Canada  N2L 3G1}
\author{Amin Eftekharian}
 \affiliation{Department of Electrical and Computer Engineering, University of Waterloo, Waterloo, Ontario, Canada N2L 3G1}
 \author{A. Hamed Majedi}
 \email{ahmajedi@uwaterloo.ca}
\affiliation{Department of Electrical and Computer Engineering,
 University of Waterloo, Waterloo, Ontario, Canada N2L 3G1}
 \affiliation{Perimeter Institute for Theoretical Physics, Waterloo, Ontario N2L 2Y5, Canada}
 \affiliation{Waterloo Institute for Nanotechnology, University of Waterloo, Waterloo, Ontario, Canada N2L 3G1}
 \author{Mohammad H. Ansari}
 \affiliation{Kavli Institute of Nanoscience, Delft University of Technology, P.O. Box 5046, 2600 GA Delft, The Netherlands}

\begin{abstract}
We study the rate of quantum phase slips in  an ultranarrow superconducting nanowire exposed to weak electromagnetic radiations. The superconductor is in the dirty limit close to the superconducting-insulating transition, where fluxoids move in strong dissipation. 
We use a  semiclassical approach and  show that external radiation  stimulates a significant enhancement in the probability of  quantum phase slips. This can help to outline a new type of detector for microwave to submillimetre radiations based on stimulated quantum phase slip phenomenon. \end{abstract}

\maketitle


\section{Introduction}
Quantum phase slip junctions are exact dual counterpart of the Josephson junctions.  Recently these junctions have been successfully realized in ultranarrow superconducting nanowires, where  quantum phase slip replaces tunneling Cooper pairs \cite{Arutyunov20081, Bezryadin}. These nanowires are nonlinear elements performing similar physics as Josephson junctions with the roles of superconducting phase  $\varphi$ and charge $q$ being interchanged  \cite{1367-2630-15-10-105017, mooijsuperconducting2006}.  This duality has been the motivation behind many of the recent applications of quantum phase slip (QPS) elements \cite{MooijSchoen, 1367-2630-7-1-219, citeulike:10580965, 4574936}. These elements have found interesting implications for fundamental metrology and information technology, for instance as photon pulse detectors, quantum current standard,  and quantum bits \cite{PhysRevB.83.174511, citeulike:10580965, 4574936, PhysRevLett.108.097001}.

In a superconducting nanowire with small cross section, the supercurrent is determined by the phase difference $\varphi$ between two ends of the nanowire from the sawtooth relation $I_s=\Phi_0 \varphi /2\pi L$ with $L$ being nanowire kinetic inductance and $\Phi_0=h/2e$.   In temperatures much lower than the superconducting critical temperature (i.e. $T\ll T_c $) quantum fluctuations may suppress the modulus of the order parameter in a region and turn it from superconductor to normal metal. This enables the superconducting phase to slip by $2n\pi$, with integer $n$, without any energy compensation.  An individual phase slip takes place in a normal core, similar to the normal core of a magnetic flux vortex, therefore we can assume the core size is roughly the coherence length $\xi$ \cite{gio, PhysRevLett.87.217003}.  QPS event takes place  for a short period of time  that is maximally  of the order of inverse of superconducting gap $h/2\Delta$.   Similar effect happens close to $T_c$ due to thermal fluctuations of the order parameter \cite{tinkham}.

In superconducting nanowire made of clean materials with low normal resistance $R$, quantum phase slips rarely take place.  To enhance the slip rate a nanowire should be made of highly disordered amorphous superconductor, which is  in the dirty limit, with large $R$ \cite{citeulike:10580965,MooijSchoen}.  There is not a well-understood theory to describe the superconductivity in near superconductor-insulator transition (SIT).  A candidate theory \cite{PhysRevB.32.5658,Sadovskii1997225} proposes superconductivity at high disorder is maintained by a fragile coherence between electron pairs, which is characterized by an anomalous binding energy. If pairs are localized, they enter an insulating state, and if condense, a coherent zero-resistance state emerges.  Based on this theory superconductor in SIT have regions  of localized BCS-condensates nearly separated in different lakes \cite{Sacepe2011jm}. The cores of QPS can coherently tunnel across superconducting regions and avoid dissipation.  This is similar to the Cooper pairs that tunnel across a Josephson junction without much dissipation \cite{PhysRevLett.46.211,Caldeira1983374}.

The voltage across the nanowire is known to be periodic in charge of the crossing fluxoid; i.e. $V=V_0\sin(2\pi q/2e)$. Individual phase slips in nanowires can be observed  when a large bias voltage is applied on the wire. Under such bias voltage, effective potential becomes a tilted washboard with more slanted slop in larger bias. Depending on temperature, there are two general scenarios  for the dynamics of a fluxoid. Close to the critical temperature $T_c$,  fluxoid particle gains energy from thermal activation and overcomes potential barrier to  slip across the wire \cite{tinkham}.  Quite differently, in low temperature $T\ll T_c$  fluxoid particle becomes frozen in a minima of the washboard potential. The minima are called  `zero-current states' where  Coulomb blockade occurs \cite{1367-2630-14-4-043014, PhysRevLett.109.187001, PhysRevB.87.144510}.    Vacuum fluctuations  help the particle to tunnel into the barrier and slips away. Coherent tunneling is possible between two zero-current states where a quantum variable (phase or charge) has minimum fluctuations. Such coherent tunnelings have been previously observed  in the superconducting-insulating transition limit \cite{1367-2630-14-4-043014, PhysRevLett.109.187001, PhysRevB.87.144510}.

Exposing nanowire to strong electromagnetic radiation produces Shapiro steps \cite{PhysRevLett.11.80,anderson64}  in the current-voltage dependence, which has been observed in reference \cite{Bae:2009hn}.  However, in many applications of superconducting nanowires, such as in qubits and photon detectors,  weak radiation is applied where phase locking cannot occur.

In this paper, we qualitatively study the effect of a weak alternating electromagnetic field on the quantum phase slip rate in ultranarrow superconducting nanowire, where the width of the nanowire is smaller than the superconducting coherence length, i.e. $r<\xi$. 
We consider the nanowire is in the insulating phase.  Our method is to map this problem into its well-studied analogue in Josephson junction in proper regime. We use the semiclassical quantum mechanical approach developed by Ivlev and Mel'nikov \cite{SovPhysJETP.63.1986, PhysRevLett.55.1614, SovPhysJetpIvelevMelnikov1985} in studying quantum tunneling in a high-frequency field to our problem. Similar to a Josephson junction under weak time-harmonic radiation \cite{PhysRevLett.53.1260,SovPhysJetpIvelevMelnikov1985}, we expect a significant enhancement in the stimulated phase slips  at zero temperature.  We show that a fluxoid   gains energy from radiation and  tunnel into the barrier more often than usual and slips away. This leads to the super-exponential enhancement  in the rate of such `stimulated quantum phase slips' (SQPs). In certain nanowires, this can  result in larger DC resistivity with minimal fluctuations in a dynamical variable.

The enhanced escape from the zero-current state stimulated by weak irradiation has significant practical importance. Since the observation of zero-current state requires a high impedance environment, our study in here will be confined to the highly dissipative cases.   Our model suggests that the QPS in certain nanowires at low temperature can significantly be amplified in microwave to THz radiation. The feasibility of this detector at the typical frequency of 0.3 THz using conventional materials will be discussed.

\section{Stimulated Quantum Phase Slips}\label{QPS_HF}

A superconducting nanowire with QPS is the dual to a Josephson junction with charge and phase (as well as current and voltage) interchanged. In Josephson junction, a Cooper pair tunneling across the junction picks up a phase $\exp(\pm i \varphi(t))$ corresponding to the superconducting phase $\varphi(t)$. This induces a coupling energy $E=E_J(1-\cos \varphi)$. The current is defined $I=(2e/\hbar ) \partial E/\partial \varphi$.

\begin{center}
\begin{figure}
\includegraphics[scale=0.5]{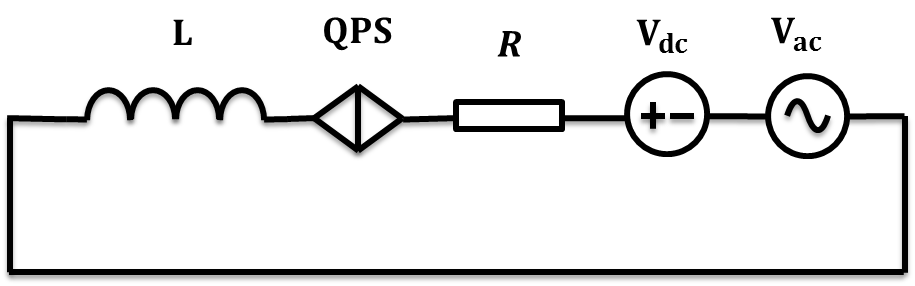}
\caption{ The schematic circuit of a QPS junction including an ideal QPS element (a superconducting nanowire), the dissipative element $R$, the bias voltage $V_{DC}$ and the driving source $V_{AC}$. $L$ denotes the dominant  kinetic inductance.}
\label{figQPScir}
\end{figure}
\end{center}

Analogously for a nanowire similar relations can be derived. A QPS fluxoid picks up a charge phase when tunneling $\exp(\pm  iQ)$  with  $Q\equiv 2 \pi q/2e$ being a dimensionless charge parameter.  Therefore the QPS energy becomes $E=E_S(1-\cos Q(t))$. The voltage is defined $V=\partial E/\partial q$. Phase slips may take place everywhere in the wire whose induced current depends on the wire inductance. Therefore a narrow superconducting nanowire can be modelled as a voltage in series with an inductance as shown in the left part of Fig. \ref{figQPScir}. In the figure, the dissipation is modelled by a resistor and the AC and DC bias voltages are sources in series with the wire. This circuit is built based on \cite{1367-2630-15-10-105017, mooijsuperconducting2006}. The inductor $L$ is the total of the kinetic inductance ($L_k$) and the geometric inductance ($L_g$) of the circuit. Since, in superconducting nanowire, the kinetic inductance is much larger than the geometric inductance, we have $L\approx L_k$.  In the circuit of Fig. \ref{figQPScir} voltage is  $ V= V_0 \sin\left( 2 \pi q/2e\right)+ L \ddot{q}+  R \dot{q}$ with the QPS and inductance energies
 \begin{eqnarray}\label{QPSjunctionDef0}
E_{S} =2eV_0/2\pi,  \quad \quad
E_L = \Phi_0^2/2 L_k.
\end{eqnarray}
where $V_0$ is the voltage scale of QPS energy $E_S$.

In the nanowire, a crossover from insulator  to  a superconducting inductor  takes place when  the inductance energy $E_L$  is increased beyond QPS energy  $E_S$.  In the superconducting phase $E_L\gg E_S$, the  fluxoid energy $E=E_S(1-\cos Q) + E_L(\phi_f)$  is dominated by the parabolic inductance energy $E_L(\phi_f)$ associated with induced phase $\phi_f$.  The parabola associated with different winding integer $n$ cross at certain energies where the small energy of QPS provide an avoided crossing gap. This makes nanowire energy to be multivalued in separated energy bands, similar to a capacitive Josephson junction.

In the opposite regime where $E_S\gg E_L$ the wire energy is dominated by QPS energy $E_S(1-\cos Q)$ which oscillates in charge.  Consider that the charge undergoes a fluctuation around its macroscopic value $q$ so that the stochastic charge is $q=q+\delta q$. Following the analogue discussion for a Josephson junction  phase, (see \cite{Ingold, Ansari, Ansari2} and references therein)  the charge fluctuation corresponds to the effective current fluctuations across the wire
\begin{equation}
\label{eq. deltaq}
\delta q(t) =  \int_0^t \delta I(t') dt'.
\end{equation}

A difference between QPS and thermally activated phase slips (TAPS) is that the dissipation in TAPS is due to  stochastic energy activation in high temperature while QPS allows tunneling between distinct zero-current states. The latter is similar to zero-voltage states in Josephson junction \cite{1367-2630-14-4-043014, PhysRevLett.109.187001, PhysRevB.87.144510}. For a nanowire in the insulating phase the quantum tunneling between zero-current states takes place without current fluctuations $\delta I(t)$. Therefore Eq. (\ref{eq. deltaq})  shows that  the charge in fluxoid  behaves semiclassically, whereas superconducting phase can be subject to large fluctuations  \cite{mooijsuperconducting2006}.

The semiclassical charge  associated with QPS fluxoid  in the nanowire depicted in Fig. \ref{figQPScir}  evolves in the following way:
\begin{equation}\label{EOMQPSinitialq}
  \frac{d^2 Q}{dt^2}+ \eta \frac{dQ}{dt}+ \omega_p^2 \left(\cos  Q  -k_{0} -  k_{1}  \cos \Omega t \right)=0,
\end{equation}
with  $\Omega$ is the frequency of the driving voltage and
\begin{equation}\label{QPSjunctionDef1}
\begin{split}
\eta  = R/ L , \quad
\omega_p = \sqrt{ 2\pi V_0/ 2 e L}, \quad
k_{i} = V_{i}/V_0.
\end{split}
\end{equation}
with the index $i=$ 0 (1) corresponds to DC (AC) voltage.  The definition of the plasma frequency $\omega_p$ is compatible with the definition based on the duality: $\hbar \omega_p =\sqrt{2 E_{S} E_L} $, which is similar to the equation of RCSJ model of Josephson junction \cite{mccumber,stewart} with high-frequency driving field. For simplicity in writing Eq. (\ref{EOMQPSinitialq}) we had shifted $Q\to Q+\pi/2$ in order to have applied and QPS voltages in phase.

Our aim is to study the possibility of utilizing  nanowire as a detector for time-harmonic radiations, therefore we restrict ourselves to weak alternating fields, i.e. $k_1 \ll 1$.  This makes our problem to be different from the physics of the Shapiro steps \cite{PhysRevLett.11.80}  where the phase of nanowires (or its dual Josephson junction) is locked to the frequency of the driving field frequency and constant voltage steps are observed. For weak time-harmonic fields the driving force is very small and the wire is in zero-current state with  $k_0<1$. The most important result is that the  smallness of $k_1$ does not necessary mean that the its effect on the charge   dynamics is small. In fact as we will show a weak time-harmonic field can significantly affect the wire by increasing the rate of QPS  (see also Appendix \ref{Sem_clas_QT}).

In the limit of weak dissipation, the tunneling rate of fluxoid can be studied using 1D quantum mechanics of the Lagrangian associated to Eq. (\ref{EOMQPSinitialq})  subject to  $R=0$. However  we are interested to study the decay of the zero-current state in the limit of strong dissipation because resistance is large in wires with QPS effects. A dual effective theory has been developed for dissipative coherent tunneling in Josephson junction by Caldeira and Leggett \cite{Caldeira1983374,PhysRevLett.46.211}, and Larkin and Ovchinnikov \cite{PhysRevB.28.6281}.  In those theories, dissipation is modelled as the coupling to bosonic degrees of freedom. The low-energy effective action of the system is derived to properly take account of the dissipation.

The semiclassical theory of  Josephson junction exposed to weak alternating current in \cite{PhysRevB.28.6281, SovPhysJetpIvelevMelnikov1985} guides us to study the charge dynamics of radiation-stimulated quantum phase slips (SQPS)  in nanowires with strong dissipation.    Effectively the evolution of semi-classical charge that tunnel across the wire is:
\begin{eqnarray}\label{EOMdualQPSwithDissipation}
\nonumber  && \frac{d^2 Q}{dt^2} +\omega_p^2 \left( \cos Q-k_0-k_1 \cos \Omega t \right) \\ &&  \nonumber \quad- 2i \pi \eta \left(\frac{k_B  T}{\hbar}\right)^2  \int_{C} dt_1 \frac{\sin [(Q(t)-Q(t_1))/2] }{\sinh^2 \left(\pi k_B T(t_1-t)/\hbar \right)} =0,\\
\end{eqnarray}
where the contour $C$ is shown in Fig. \ref{timecontour} and the principle value of the integral is implied (for general discussion of the method, see  Appendix \ref{Sem_clas_QT}).

In the limit of our interest, the nanowire is strongly dissipative $\eta \gg \omega_p$. For semi-classical description to be valid, it is required that $E_S \gg \hbar \Omega$ \cite{PhysRevB.28.6281, SovPhysJetpIvelevMelnikov1985}. Also the applied DC voltage is close to $V_0$, i.e. $V_0-V_{DC} \ll V_0$, therefore the term with second derivative in Eq. (\ref{EOMdualQPSwithDissipation}) can be omitted. Also, we can assume that the exchange of energy between the wire and its environment takes place in the shortest time, thus  the argument of $\sinh$ in the denominator of Eq. (\ref{EOMdualQPSwithDissipation}) can be replaced by its lowest order $\sinh x\approx x$. Technical analysis of this integration over the contour $C$ shows that the integral tends to zero except that at the singularity $t=t_1$ where its proper residue must be counted (see Eq. (18) in Ref. \cite{SovPhysJetpIvelevMelnikov1985}). Therefore, in the lack of alternating radiation  Eq.(\ref{EOMdualQPSwithDissipation}) in the regime of interest effectively reads: $-\eta dQ/dt+ \omega_p^2 \left( \cos Q-1 \right)  =0$, which has the following solution
\begin{equation}\label{eq.sol}
Q(t)=i \ln \frac{t-i\tau_s}{t+i\tau_s} , \quad \tau_s=\frac{\eta}{\omega_p^2},
\end{equation}
with $\tau_s$ being the time of under-barrier motion.

In a system described by the classical action $S=-i\int_C  \mathcal{L} dt$
with $\mathcal{L}$ being  Lagrangian,  the probability of quasiclassical tunneling is $\Gamma=\exp(-S)$. Regarding the alternating voltage being a small perturbation $k_1\ll 1$, we can rewrite the action in the form of $S=S_0+S_1$ with $S_0$ being the action in the lack of alternating field and $S_1$ is  linear in $V_{AC}$, \cite{kagan1992quantum}.

Let us assume the QPS probability in a nanowire with strong dissipation and DC voltage $V_{DC}$ about $V_0$ is denoted as $\Gamma_0$.  Above results  easily show that in the presence of a weak high frequency radiation hitting the nanowire, the probability of QPS in the wire will change from  $\Gamma_0$ to $\Gamma$ in the following form:
\begin{equation}\label{tunnelingprobACwithdissipation}
\Gamma(V_{AC},\Omega)= \Gamma_0 \exp\left[ \frac{4e V_{AC} }{ \hbar \Omega}  \sinh (\Omega \tau_s) \right],
\end{equation}

Eq.(\ref{tunnelingprobACwithdissipation}) is the main result in this paper. It indicates that an alternating driving field with certain frequency and voltage can trigger occurance of a large number of QPS's in a proper nanowire at low temperature. For instance if a wire with dissipation factor $\eta/ \omega_p = 10$ is driven by a weak time-harmonic radiation of the realtive amplitude $4 e V_{AC}/\hbar\omega_p= 5\times 10^{-4} $ and   frequency $\Omega =\omega_p$, the QPS rate increases by a factor of about   $250$ times. A more careful analysis shows that this result is valid for nanowire temperature $T<T_0$ with $T_0$ being the crossover temperature between quantum and thermal activation regimes $T_0= \sqrt{(1-k_0)/2}\  (\hbar \omega_p)^2 /\pi k_B \eta$, \cite{kagan1992quantum}.

From Eq.(\ref{tunnelingprobACwithdissipation}) one can see that the bigger the normal resistance $R$ is, the larger the rate of SQPS becomes.   Intuitively this can be understood from the definition of underbarrier time $\tau_s$ in Eq. (\ref{eq.sol}). Upon increasing dissipation the under-barrier time grows larger.  According to the semiclassical description of quantum tunneling, see Appendix (\ref{Sem_clas_QT}), during quantum tunneling time parameter becomes imaginary. This changes the bounded function of alternating potential in eq. (\ref{EOMQPSinitialq})  into the unbounded function $\cosh \Omega \tau$. The longer a fluxoid stays under the barrier, the more energy it absorbs from the alternating potential and this causes stimulation of quantum phase slips.

The QPS rate in the lack of an time-harmonic drive has been estimated by Mooij and Harmans in \cite{1367-2630-7-1-219} to be nearly $\Gamma_0 \approx E_{S}/\hbar$. This in addition to substituting Eqs. (\ref{QPSjunctionDef1}) and (\ref{QPSjunctionDef0}) simplifies Eq.  (\ref{tunnelingprobACwithdissipation}) into:
\begin{equation}\label{tunnelingprobACwithdissipationEQPS}
\Gamma(V_{AC}, \Omega)= (V_0 /\Phi_0) \exp\left[ \frac{4e V_{AC} }{ \hbar \Omega} \sinh \big(\frac{e  R \Omega}{\pi V_0}\big)\right],
\end{equation}
Given that the QPS rate in the absence of time-harmonic radiation is proportional to $V_0$,  one of the features of Eq. (\ref{tunnelingprobACwithdissipationEQPS}) is that the super-exponential enhancement of QPS rate is inversely proportional to the rate $\Gamma_0$, thus for small rate $\Gamma_0$ the exponential enhancement of QPS in the presence of high-frequency field is more significant. This enhancement is only due to the stimulated excitation in the zero-current states in the presence of external drive.

In the absence of the $V_{DC}$, there is no tilt in potential and the rate of fluxoid crossing to right or left are equal. As a result  the average current becomes zero $(\bar{I}=0$). However, in the presence of a positive value for DC voltage bias, the average current is given by:
\begin{equation}
\bar{I}=2e \left( \Gamma_{\rightarrow}-\Gamma_{\leftarrow} \right)
\end{equation}
with  $\Gamma_{\rightarrow}$ ($\Gamma_{\leftarrow}$) the rate of crossing to the right (left) where the potential barrier decreases (increases). In the case the bias voltage is close to the critical voltage ($V_0$),   $\Gamma_{\leftarrow}$  the crossing $\Gamma_{\rightarrow}$ dominantly exceeds that of the opposite direction. Hence $\Gamma= \Gamma_{\rightarrow}$ and
\begin{equation}\label{averagIQPS}
I= 2e  \Gamma(V_{AC}, \Omega),
\end{equation}
where $\Gamma(V_{AC}, \Omega)$ is given by Eq. (\ref{tunnelingprobACwithdissipation}). According to Eq. (\ref{averagIQPS}), the influence of high-frequency weak  irradiation on  superconducting nanowire biasd by the DC voltage $V_0-V_{DC}\ll V_0$  is observable by measuring the crossing current.

The quality factor in nanowire  $Q_{S}$ is defined as:
\begin{equation}
Q_{S}= \frac{\omega_p}{\eta}.
\end{equation}
In low quality factor QPS nanowire the dissipation is strong. The larger $\eta$ leads to longer under-barrier time and consequently the enhancement of SQPS rate exponentially increases. The increase in the under-barrier motion due to higher dissipation can be seen in Fig. \ref{multiphoton_tunneling} . Therefore, we expect that a low-$Q_S$ nanowire to be a better candidate for observing tunneling enhancement.

A comment on the range of validity of the method we used in this section is in order. As it is seen from Eq. (\ref{tunnelingprobACwithdissipation}), the enhancement in the tunneling probability for $\Omega \eta \gg \omega_p^2$
is itself an exponentially large factor ($\sim e V_{AC}( \hbar\Omega)^{-1}\exp (\Omega \eta/\omega_p^2) $). This indicates that the range of the validity of the semi-classical approach in this case is limited to $e V_{AC} \sim (\hbar\Omega) \exp (-\Omega \eta/\omega_p^2)$. Beyond this, higher order correction in terms of $V_{AC}$ to the calculations is required \cite{kamenev2011field}.

An alternative approach in studying the QPS rate in superconducting nanowires under high-frequency radiation would be to use the the effective action method developed by Golubev and Zaikin  \cite{1999cond.mat.11314Z,PhysRevB.64.014504,PhysRevLett.78.1552,Arutyunov20081} in a non-equilibrium setting. There are some challenges associated with this approach that are studied in \cite{phdthesisamirjafarisalim}.

\section{Proposal for QPS-based Energy-resolving High-frequency Radiation Detector}\label{HF_QPSdetector}

The exponential enhancement of the  probability  of the quantum tunneling observed in  Eq. (\ref{tunnelingprobACwithdissipation}) can be exploited in designing detectors of microwave to THz radiation that are capable of determining the frequency of the incoming energy. In this section we introduce a new type of high-frequency detectors based on enhancement of QPS phenomenon in superconducting nanowires. In this paper, the working principle of this type of detector will be discussed and  much important engineering details like impedance matching will not be addressed.



\begin{figure}
\begin{center}
\includegraphics[scale=0.35]{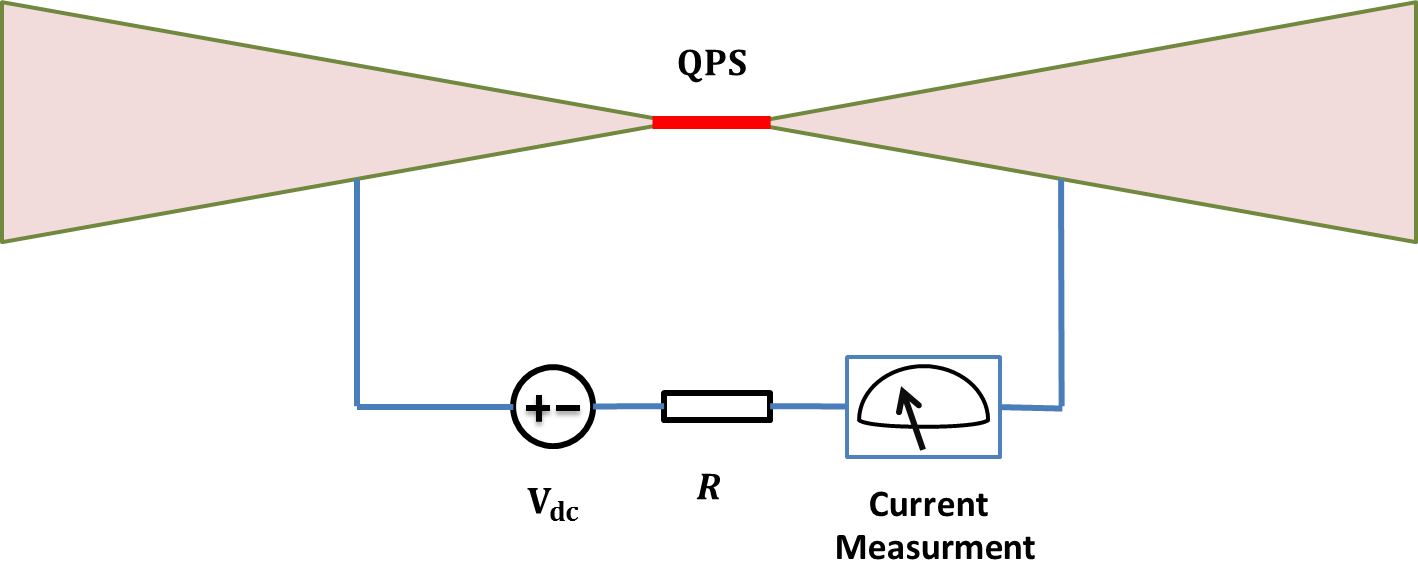}
\caption[The schematic of a QPS high-frequency detector is shown. The red segment in the middle, is the superconducting nanowire. A broadband bow-tie antenna collects the high-frequency field.]{ The schematic of a QPS high-frequency detector is shown. The red segment in the middle, is the superconducting nanowire. A broadband bow-tie antenna collects the high-frequency field. The resistance $R$ adds dissipation to lower the quality factor of the QPS junction.}
\label{Bowtieantenna}
\end{center}
\end{figure}

Our proposed detector is made of a low quality factor, i.e. low $Q_S$, QPS junction that is voltage biased close to the critical voltage. An antenna is the source of the high-frequency voltage and is placed right across the superconducting nanowire. The current in the loop is measured constantly, the change in the current and the amplitude determines the presence of the detected radiation. A schematic of this system is shown in Fig. (\ref{Bowtieantenna}).


The superconducting nanowire is placed in the gap between two parts of the antenna. This will guarantee that the maximum $V_{AC}$ is induced along the nanowire. Other elements like the resistance and the voltage-bias source are placed outside of the antenna in the loop.

Presence of the radiation results in the decay of the zero-current state of the QPS junction which causes a change in the current of the circuit. Depending on the design parameters, the detection of the change in the current might be hard to achieve. A lock-in amplifier or a SQUID can be used for current monitoring in case the current change is difficult to be monitored with conventional methods.

The resistance $R$ plays the important role of reducing quality factor ($Q$) of the QPS junction. Its value is chosen such that the required enhancement in Eq. (\ref{tunnelingprobACwithdissipation}) is achieved, which depends on other parameters of the system.

\subsection{Design Parameters}
As an example, in this section, we investigate design parameters for a $\Omega/2\pi=0.3$ THz detector. We will study   different superconducting materials to evaluate the applicability of them in our design. Based on these properties, parameters of the detector can be estimated.

We assume that the QPS energy is related to the QPS rate according to $E_{S}=\hbar \Gamma_{QPS} $ \cite{1367-2630-7-1-219}. The QPS rate from the Golubev-Zaikin theory  \cite{1999cond.mat.11314Z,PhysRevB.64.014504,PhysRevLett.78.1552,Arutyunov20081} is given by
\begin{equation}\label{QPS_rate_detector}
\Gamma_{QPS}= c_1 \frac{\Delta}{\hbar}\frac{R_q}{R_n} \frac{X^2}{\xi^2} \exp \left(- 0.3 c_2 \frac{R_q}{R_n}\frac{X}{\xi}  \right),
\end{equation}
where $R_n$ is the normal resistance per unit length of the superconducting nanowires. The two constants $c_1$ and $c_2$ account for uncertainties in derivation of Eq. (\ref{QPS_rate_detector}) which are of order one. We set $c_1=c_2=1$.
Although Eq. (\ref{QPS_rate_detector}) is given by  Golubev-Zaikin theory, the factor $0.3$ in the exponent is adopted from the fit of experimental data to the Giordano model in the work of \cite{PhysRevLett.87.217003,1367-2630-7-1-219}.

In order to choose the appropriate material and parameters for the detector, we study properties of four different materials NbSi, InO$_x$, NbN and Ti. Properties of these materials are listed in Table \ref{materialproperties}. The coherence length $\xi$ for superconducting nanowire is related to the bulk parameter through
\begin{equation}
\xi \sim 0.85 \sqrt{\xi_{\mbox{bulk}} l_{0}},
\end{equation}
where, $l_{0}$ is the mean free path of the electrons. narrow superconducting nanowires are always in the dirty limit.
\begin{table}[h]
 \centering
 \caption{Material properties of the superconducting nanowires used for simulations. }
\begin{tabular}{c c c}
\\
\hline
Material & $\Delta$ [meV] & $\xi$ [nm]  \\ \hline
NbSi &0.18  & 15 \\
 InO$_x$ &  0.41 & 20\\
  NbN & 1.6  & 4 \\
   Ti & 0.06 & 80\\  \hline
\end{tabular}
\\
 \smallskip
\footnotesize {The data for NbSi, InO$_x$, NbN and Ti are adopted from \cite{PhysRevB.87.144510}, \cite{citeulike:10580965}, \cite{2013arXiv1305.6692P} and \cite{PhysRevLett.109.187001} respectively.}
\label{materialproperties}
\end{table}

\begin{figure}
\begin{center}
\includegraphics[scale=0.6]{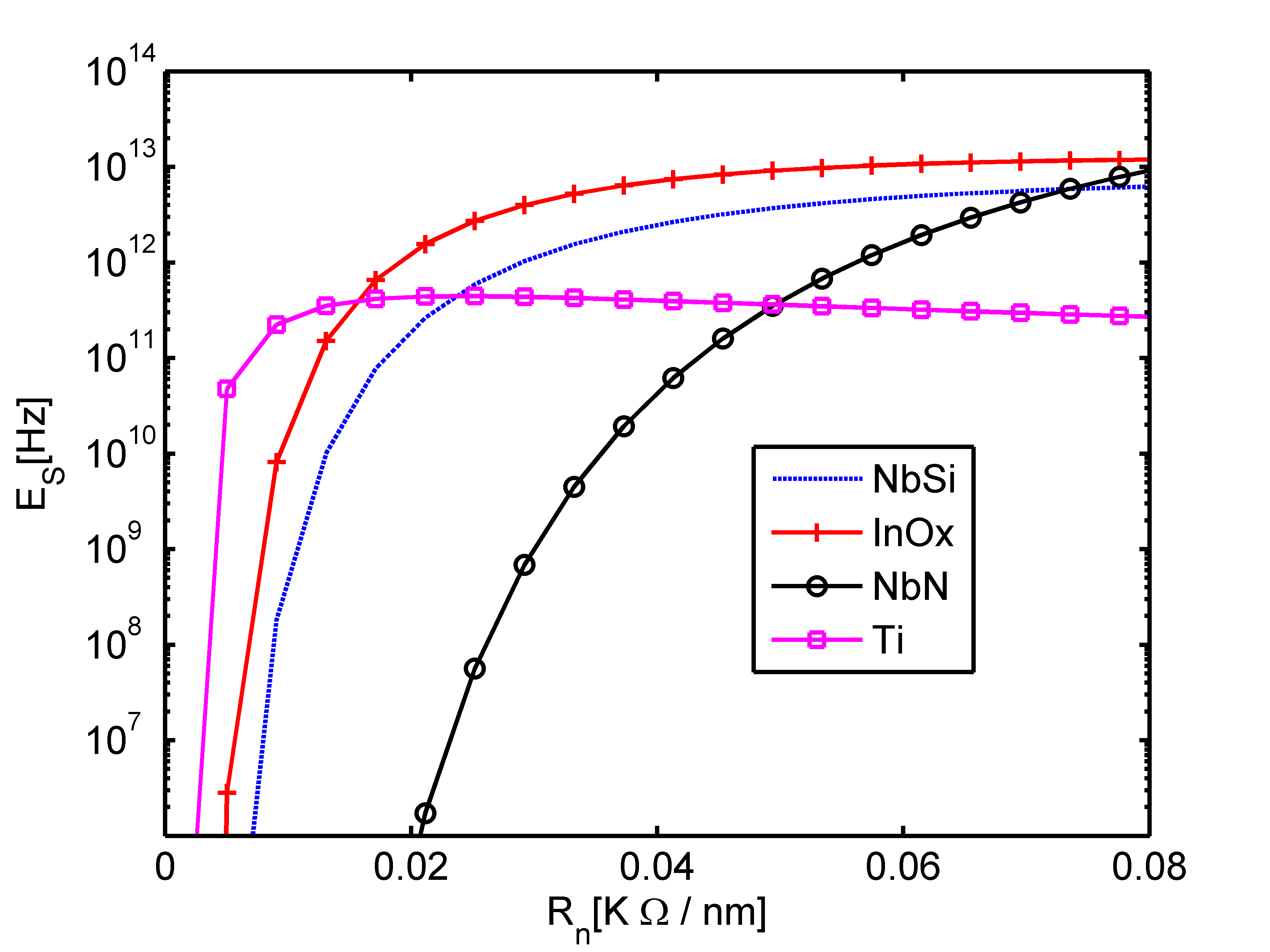}
\caption{The QPS energy as a function of the normal resistance per length for four different materials NbSi, InO$_x$, NbN and Ti is shown. The length of the nanowire is $X=2 \mu$m. Parameters are listed in Table \ref{materialproperties}. }
\label{EQPS}
\end{center}
\end{figure}

 Fig.\ref{EQPS} shows $E_{S}$ for four different materials as a function of normal resistance per  length. The resistance per length  $R_n$ determines the cross-section area of the superconducting nanowire. Since $R_n$ is inversely proportional to the cross-section area of the nanowire, the higher $R_n$ indicates thinner nanowires that support smaller current which makes the operation more difficult.

Another important energy scale in QPS junctions is the kinetic inductive energy $E_L$ which plays an important role in the dynamics. The kinetic inductive energy is given by
\begin{equation}\label{KinInductanceenergy}
E_L=\frac{\Phi_0^2}{2 L_k},
\end{equation}
where the kinetic inductance is found from
\begin{equation}\label{KinInductance}
L_k=\frac{\hbar R_N}{\pi \Delta}.
\end{equation}
In Eq. (\ref{KinInductance}), $R_N$ is the total normal state resistance of the superconducting nanowire which is given by $R_N=X R_n$, where $X$ is the length of the superconducting nanowire. In Eq. (\ref{KinInductanceenergy}), the geometric inductance and external inductance are assumed to be much smaller than the kinematic inductance $L_k$ of the superconducting nanowire.

\begin{figure}
\begin{center}
\includegraphics[scale=0.6]{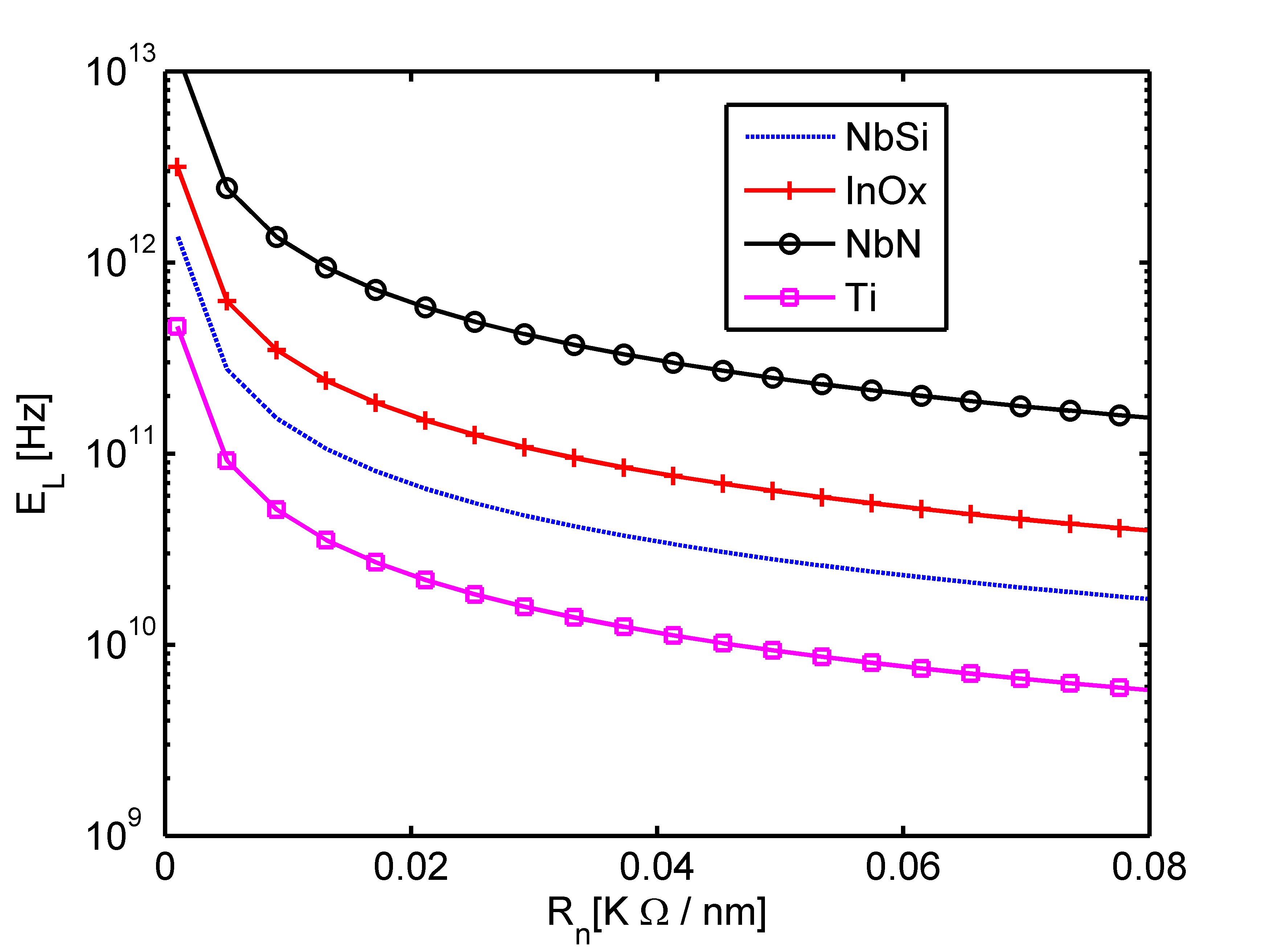}
\caption{ The inductive kinetic energy as a function of the normal resistance per length for four different materials NbSi, InO$_x$, NbN and Ti is shown. The length of the nanowire is $X=2 \mu$m. Parameters are listed in Table \ref{materialproperties}.  }
\label{EL}
\end{center}
\end{figure}

\begin{figure}
\begin{center}
\includegraphics[scale=0.6]{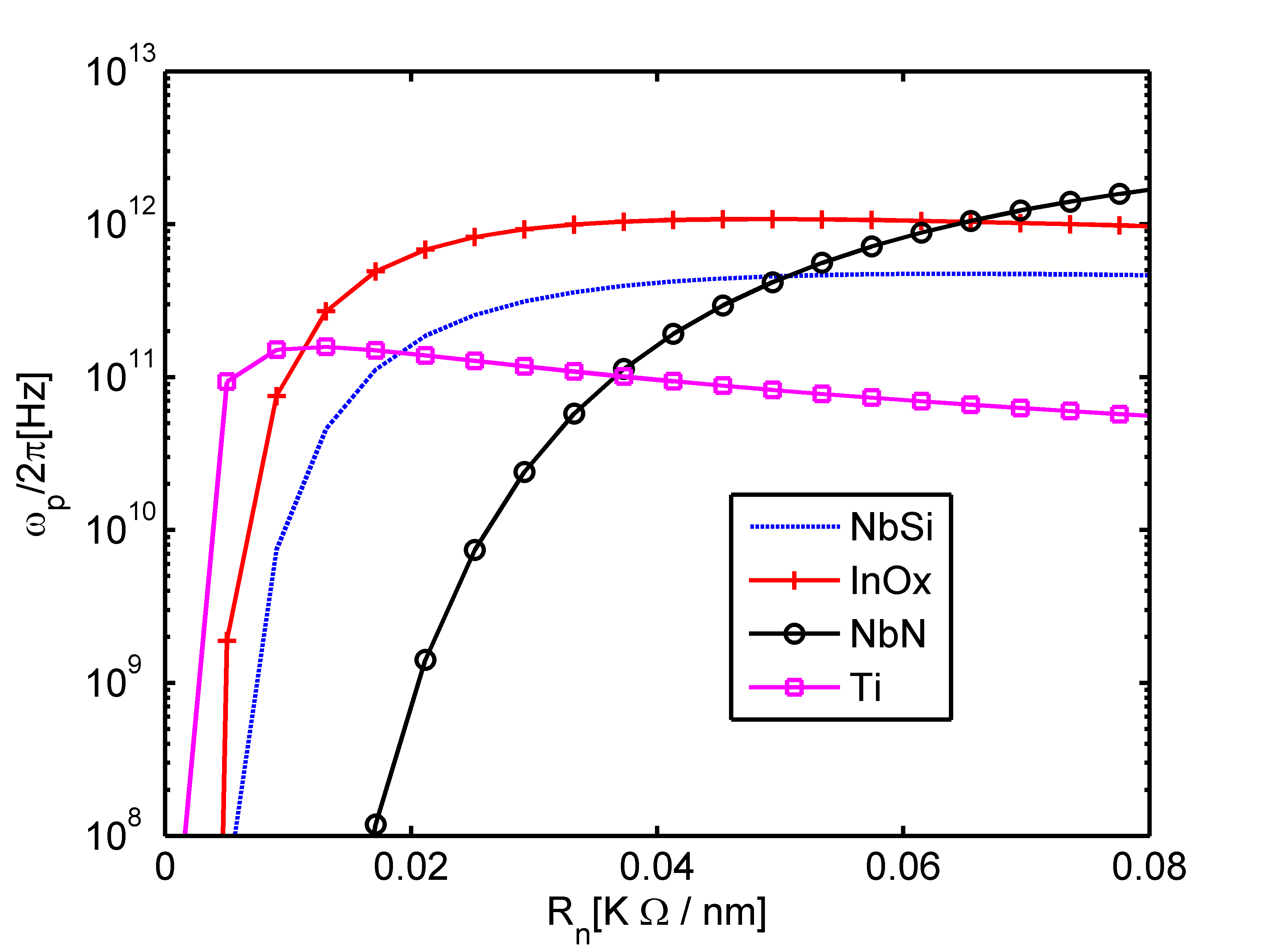}
\caption{ The plasma frequency $\omega_{p}/2\pi$ as a function of the normal resistance per length for four different materials NbSi, InO$_x$, NbN and Ti is shown. The $R_n$ determines the dimensions of the superconducting nano wire. The length of the nanowire is $X=2 \mu$m. Parameters are listed in Table \ref{materialproperties}.  }
\label{omegapl}
\end{center}
\end{figure}

In Fig. \ref{EL}  and Fig. \ref{omegapl} the inductive kinetic $E_L$ energy and the plasma frequency $\omega_p$ of the four nanowire as a function of normal resistance per length are shown.

\begin{figure}
\begin{center}
\includegraphics[scale=0.6]{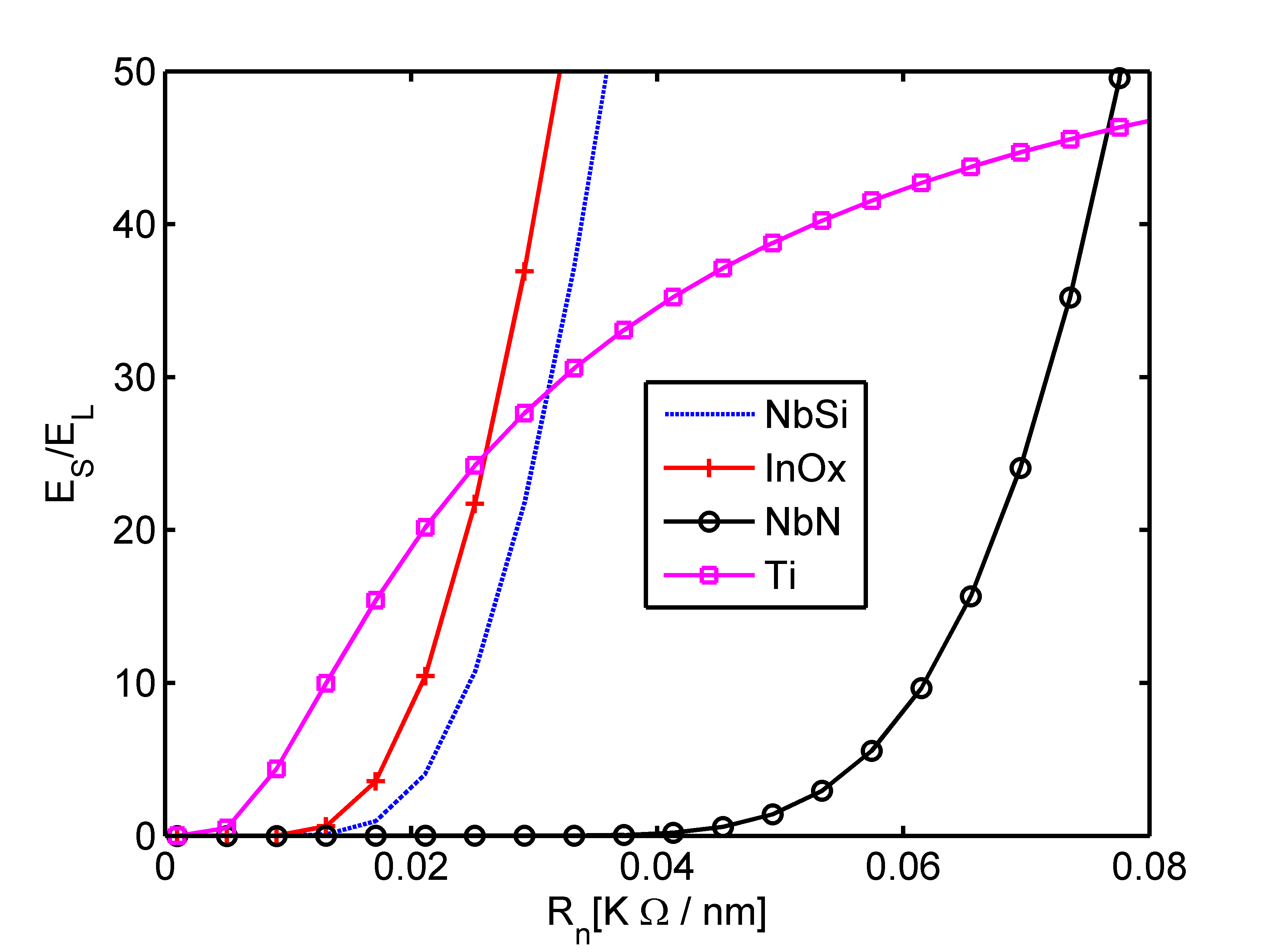}
\caption{ The ratio of $E_{S}/E_{L}$ as a function of the normal resistance per length for four different materials is shown. For the charge number to be the good quantum number, it is required that $E_{S}> 4E_{L}$. }
\label{EQPSoverEL}
\end{center}
\end{figure}

Since the superconducting nanowire is intended to be working in the regime where charge is a good quantum number; this requires that at least $E_{S} > 4 E_{L}$. The ratio of $E_{S}/E_{L}$ is shown in Fig. \ref{EQPSoverEL}. The acceptable region of parameters is anywhere above 4.

Arbitrarily, we choose the length of the superconducting nanowire to be $X=2\mu$m. To satisfy the conditions of the semiclassical approach of the previous section, i.e., $E_S \gg \hbar \Omega$, we choose  $E_{S}=3$ THz for the operating frequency of $\Omega/ 2\pi = 0.3$ THz. This leads to the critical voltage of
$V_0=\frac{ 2\pi}{2e} E_{S}= 39 \mbox{mV}$.

Assuming the induced alternating voltage collected by the antenna has the amplitude of $V_{AC}=100$ nV, then $4eV_{AC}/\hbar \Omega \sim 10^{-3}$. Therefore, from Eq. (\ref{tunnelingprobACwithdissipationEQPS}), to have a significant enhancement it is necessary to have
\begin{equation}\label{enhancementcondition}
\frac{\Omega \eta}{\omega_p ^2 } \gg \sinh^{-1} (10^{3}) \approx 8,
\end{equation}
where the dissipation $\eta$ with dimension radian per second is defined in Eq. (\ref{QPSjunctionDef1}). Fig. \ref{EQPS} shows that InO$_x$ could be a possible candidate. In order to obtain $E_{S}=3$ THz, from  Figs. \ref{EQPS}, \ref{EL}, \ref{omegapl} and \ref{EQPSoverEL} we obtain the following parameters for InO$_x$ nanowire: $R_n \approx 0.025$ K Ohm/nm, $E_L\approx 120$ GHz, $\omega_p/2\pi\approx 0.84$ THz, and $E_S/E_L \approx$ 24.
Using these parameters in Eq. (\ref{enhancementcondition}), in order to have significant enhancement the total resistance of the circuit needs to be much larger than
$490  \,\, \mbox{K Ohm}$. For an on chip resistance, NiCr thin-film resistors can be used.

One of the significant advantages of the proposed detector would be the simplicity of the fabrication using one-layer lithography. The superconducting nanowire and the antenna can be fabricated lithographically on the same substrate. Avoiding shunting parasitic capacitances might be challenging that requires extra attention.


The width of the QPS element is in the order of $10$ nm to $20$ nm, therefore, a large number of them can be fabricated in parallel which makes them good candidates for applications that require many elements like imaging or for higher detection and coupling efficiency.

In the proposed method the detector is made of narrow nanowires with width smaller than the coherence length, therefore the presence of vortices can be ignored. This is because structures smaller than $4.4 \xi$ can not support vortices. The absence of vortices might improve the noise performance of the proposed detector.


\section{Concluding Remarks}

We studied the stimulation effect of a weak high-frequency field on the zero-current state tunneling of fluxoid particle. The approach  chosen was to use the duality transformation between Josephson junction and a QPS junction to map the dynamics of QPS charge  in a circuit model. Then we studied the effect of high-frequency alternating field on the coherent tunneling rate. The similar problem has been studied for the case of Josephson junction using semiclassical physics \cite{SovPhysJETP.63.1986, PhysRevLett.55.1614, SovPhysJetpIvelevMelnikov1985} which we adopted for the case of QPS junction. We observed that in a strongly dissipative superconducting ultranarrow nanowire, a high frequency field can enhance the probability of quantum tunneling super-exponentially.  Interestingly we find that  the enhancement of SQPS rate is more pronounced in wires with small non-stimulated QPS. This result will help to predict that quantum phase slip qubits should be better-working in the presence of weak driving field.

The rate enhancement and its driving frequency dependence can be exploited in designing energy-resolving high-frequency detectors. We outlined a new type of high-frequency detectors based on the QPS phenomenon in superconducting nanowires. The basic physics and design for such a detector was introduced. We then investigated the possibility of such realization using the materials used in studying of QPS in superconducting nanowires. It was shown that the theoretical restrictions can be met by choosing correct design parameters. 

\begin{acknowledgments}
This work is supported by the NSERC Discovery Grant. This research is also sponsored by CryptoWorks21, an NSERC funded collaborative research and training experience program. AJS and MHA thank Frank K. Wilhelm-Mauch for fruitful discussions.
\end{acknowledgments}

\appendix

\renewcommand\thesection{\Alph{section}}
\renewcommand{\thefigure}{\thesection.\arabic{figure}}

\section{\label{Sem_clas_QT} Brief Review of the Quantum Tunneling in Time-dependent Potentials}

In this appendix, we review the effect of a high-frequency field on quantum tunneling in the semi-classical description\cite{kamenev2011field}.
The approach will be based on the method developed in \cite{ SovPhysJetpIvelevMelnikov1985,PhysRevLett.55.1614,SovPhysJETP.63.1986}. First, a brief introduction is given to the semi-classical approach to quantum mechanics and quantum tunneling.

\begin{figure}
\centering
\includegraphics[scale=0.5]{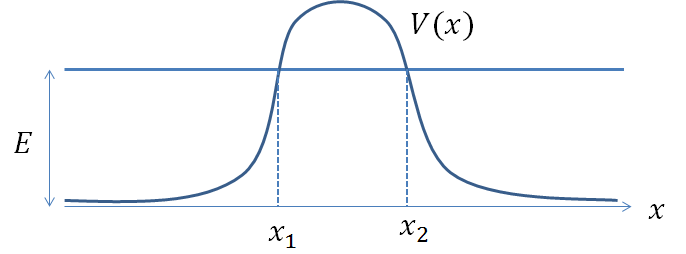}
\caption{ Potential barrier for a particle moving from left to right with energy $E$. Classical turning points are indicated by $x_1$ and $x_2$. According to Eq. (\ref{timeofpointparticle}), for $x_1<x<x_2$, the time becomes complex.}\label{tunnelingpotential}
\end{figure}

The semi-classical description is obtained from the stationary path approximation of the Feynman path integral approach to quantum mechanics. The stationary path of a Feynman path integral which is obtained from the variation of the action yields the Newtonian equation of motion (EOM):
\begin{equation}
\delta S=0 \longrightarrow \mbox{EOM}
\end{equation}
This relation is familiar in classical mechanics  for energetically allowed region; however, the natural question that arises is that: is this method applicable to energetically forbidden regions like in quantum tunneling? The answer is ``yes''; however, it requires allowing the time to acquire an imaginary part \cite{landau1977quantum}. To see this, let's consider the action of a point particle with mass $m$ in a potential $V(x)$. The action can be written as:
\begin{equation}\label{actiontheoreticalbackground}
S=\int dt \left\{ \, \frac{m}{2} \left( \frac{d x}{dt} \right)^2-V(x)+E \right\}.
\end{equation}
The equation of motion is found to be
\begin{equation}\label{EOMtheoreticalbackground}
 m \frac{d^2 x}{dt^2}+ \frac{dV(x)}{dx}=0,
\end{equation}
and the total energy is given by
\begin{equation}\label{energyofpointparticle}
E=\frac{p^2}{2m}+V(x),
\end{equation}
where the  momentum is defined as $p(t)=m dx/dt$. By integrating Eq. (\ref{energyofpointparticle}), the required time $t$ for the particle to reach infinity from point $x$ is given by
\begin{equation}\label{timeofpointparticle}
t(x)=\int_{x}^{\infty} \frac{dx' \sqrt{m}}{\sqrt{2 \left( E-V(x')   \right)}}.
\end{equation}
From Eq. (\ref{timeofpointparticle}), it is seen that as long as $E>V(x)$ the time remains real, but for $E<V(x)$ it acquires an imaginary part and becomes complex. Therefore, by allowing complex time, classically forbidden regions can be studied in the semi-classical approach. The tunneling of a point particle with energy $E$ coming from left from the potential $V(x)$ is shown in Fig. \ref{tunnelingpotential}. For $x>x_2$ time is real, because $E>V(x)$; however, for $ x_1<x<x_2$, the time goes in the imaginary direction. For $x<x_1$ time becomes complex, $t+i \tau_0$, where the constant imaginary part is
\begin{equation}\label{timeunderbarrier}
\tau_0=\int_{x_1}^{x_2} \frac{dx \sqrt{m}}{\sqrt{2\left( V(x)-E \right) }}.
\end{equation}

\begin{figure}
\centering
\includegraphics[scale=0.5]{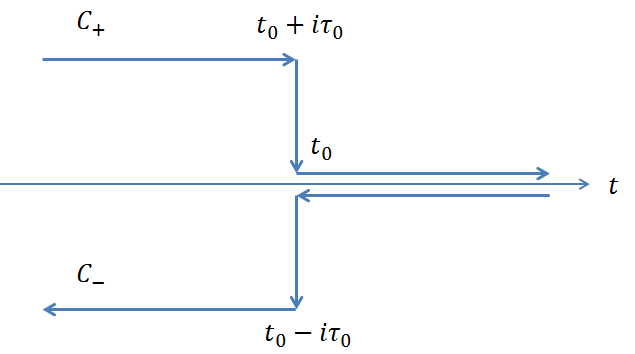}
\caption{ The integration contour for the quantum tunneling probability. The vertical sections correspond to the underbarrier motion.}\label{timecontour}
\end{figure}

Therefore, for a tunneling path of a particle moving from left to right, the time evolution is depicted in Fig. \ref{timecontour} by contour $C_+$.

According to the semi-classical description, the tunneling amplitude is found by calculating the action Eq. (\ref{actiontheoreticalbackground}) along contour $C_+$. In order to find the tunneling probability amplitude the contour $C_-$ needs to be added, where the property $x(t^*)=x^*(t)$ has been used. Therefore, the tunneling probability with exponential accuracy is given by
\begin{equation}\label{tunnelingprobtheoreticalbackground}
\begin{split}
\Gamma & \approx \exp\left(- S_0  \right), \\
S_0 & =- i \int_{C_-+C_+} dt \left[ \frac{m}{2} \left( \frac{d x}{dt} \right)^2-V(x)+E \right],
\end{split}
\end{equation}
where $x(t)$ is the solution to the classical equation of motion, i.e. Eq. (\ref{EOMtheoreticalbackground}), along the contour.

The horizontal segments of $C_+$ and $C_-$ cancel each other and only  vertical segments corresponding to the under barrier motion survive. Using Eq. (\ref{EOMtheoreticalbackground}) in the exponent of Eq. (\ref{tunnelingprobtheoreticalbackground}) we get
\begin{equation}\label{tunnelingprobtheoreticalbackgroundWKB}
\begin{split}
\Gamma & \approx \exp\left( i \int_{i\tau_0}^{-i\tau_0}  dt \, m \dot{x}^2 \right)= \exp\left( 2 i m \int_{x_1}^{x_2}  dx \dot{x} \right)\\ &=\exp\left( - 2  \int_{x_1}^{x_2}  dx \sqrt{2m(V(x)-E)}\right),
\end{split}
\end{equation}
which is the well-known WKB result in quantum mechanics \cite{landau1977quantum}.

If the system is in thermodynamic equilibrium before the tunneling starts, then the tunneling probability needs to be statistically averaged over  $E$
\begin{equation}\label{QTGibbs}
\langle \Gamma \rangle= \int dE \exp \left[ -\frac{E}{k_B T} -S(E) \right],
\end{equation}
where $E$ is given by Eq. (\ref{tunnelingprobtheoreticalbackground}). The largest probability of tunneling occurs for energies that  minimizes the exponent in Eq. (\ref{QTGibbs}) and is given by
\begin{equation}\label{Gibbsexponentderivative}
\frac{\partial S(E)}{\partial E}= -\frac{1}{k_B T}.
\end{equation}
The action for the underbarrier motion is given by
\begin{equation}
S(E)= 2  \int_{x_1}^{x_2}  dx \sqrt{2m(V(x)-E)},
\end{equation}
and the energy derivative of the action yields
\begin{equation}\label{actionderivative}
\frac{\partial S(E)}{\partial E}= -2  \int_{x_1}^{x_2} dx \frac{\sqrt{m}}{\sqrt{2(V(x)-E)}}=-2\tau_0
\end{equation}
where $\tau_0$ is the time of the under barrier motion given by Eq. (\ref{timeunderbarrier}).
 Comparing Eqs. (\ref{Gibbsexponentderivative}) and (\ref{actionderivative}) reveals that
\begin{equation}\label{underbarrier_time_energy}
\tau_0=\frac{1}{2k_B T}.
\end{equation}
Eq. (\ref{actionderivative}) determines the energy of the tunneling particle. Therefore, in equilibrium, the probability of tunneling through the barrier is given by Eq. (\ref{tunnelingprobtheoreticalbackground}) for real trajectories that satisfy Eq. (\ref{underbarrier_time_energy}). The real trajectories condition comes from the analysis that shows that the time-averaged probability for semiclassical processes is entirely determined by real trajectories \cite{SovPhysJETP.63.1986}.

 The semi-classical method  in which time can take on complex values is suitable for generalization to include tunneling from time dependent potentials. tunneling from periodically modulated potential barriers is the most common application of this method and since in this paper we are interested in sinusoidal alternating field we restrict this section of this review to potentials of the form \cite{SovPhysJETP.63.1986, PhysRevLett.55.1614, SovPhysJetpIvelevMelnikov1985, Caldeira1983374}
\begin{equation}
U(x,t) = V(x)+ \mathcal{E} x \cos \Omega t.
\end{equation}
According to the semiclassical description, the linear in the field-strength correction to the tunneling  probability is then given by
\begin{equation}\label{tunnelingprobhighfreqtheoreticalbackground}
S_1= - i \mathcal{E} \int_{C_-+C_+} dt \,  x (t) \cos \Omega t ,
\end{equation}
where $x(t)$ is the solution of the unperturbed equation of motion Eq.(\ref{EOMtheoreticalbackground}) along the contour. During the under-barrier motion in the vertical segment of Fig. \ref{timecontour}, time is imaginary $t=i\tau$ and therefore the equation of motion becomes
\begin{equation}
m \frac{d^2 x}{d \tau^2}- \frac{dV(x)}{dx}=0,
\end{equation}
which in comparison to Eq. (\ref{EOMtheoreticalbackground}) can be interpreted as the classical equation of motion in the inverted potential.

The contour in Eq. (\ref{tunnelingprobhighfreqtheoreticalbackground}) may be shifted to entails the singularities of the integrand. This enables calculating the integral based on the singularities of the $x(\tau)$. Therefore, the general trend of the $A_1$ in Eq. (\ref{tunnelingprobhighfreqtheoreticalbackground}) depends on the specific form of the potential. In some cases $S_1$ can be exponentially large, which is the case we encountered in this study. In order to have exponential enhancement, it is necessary for the function defined as
\begin{equation}
h(x)= \sqrt{E- V(x)},
\end{equation}
to have singularities off the real axis in the $x$ plane \cite{kagan1992quantum}. Assuming $V(x)$ has singularities of the form
\begin{equation}
V(x)\approx  \left\{
        \begin{array}{ll}
            \kappa (x-x_s)^\alpha,\quad \alpha<0, \quad  x\rightarrow x_s, \\
             \kappa x^\alpha, \quad \alpha>0,  \quad x\rightarrow \infty,
        \end{array}
    \right.
\end{equation}
then, the solution to Eq. (\ref{EOMtheoreticalbackground}) near $x_s$ is of the form
\begin{equation}
x(t)=x_s+\left[- \frac{\kappa}{2m}(2-\alpha)^2(t-t_s)^2    \right] ^\frac{1}{2-\alpha},
\end{equation}
where $t_s$ is the complex time that takes going from $x_2$ in Fig. \ref{tunnelingpotential} to $x_s$ given by
\begin{equation}
t_s=\int_{x_2}^{x_s}  \frac{dx' \sqrt{m}}{\sqrt{2 \left( E-V(x')   \right)}}.
\end{equation}
By comparing to Eq. (\ref{timeunderbarrier}), the time of the underbarrier motion $\tau_0$ has the same order of the magnitude as
\begin{equation}
\tau_s=\mbox{Im} \, t_s.
\end{equation}
but for analytical potentials always $\tau_S <\tau_0$.
For $\Omega \tau_s \gg 1 $, the main contribution to the integral in Eq. (\ref{tunnelingprobhighfreqtheoreticalbackground}) comes from branch-cut section in the vicinity of the singular points $\tau_s$ and $\tau_s^*$. Therefore, the transition probability is given by
\begin{equation}\label{t_dep_tunneling_enhancement_app}
\Gamma(\mathcal{E},t)=\Gamma_0 \exp \left( a_1\cos(\Omega t)  \right),
\end{equation}
where
\begin{equation}
a_1= \frac{2\pi \mathcal{E}}{\Omega} \left| \Gamma_{E} \left( \frac{2}{\alpha-2}\right) \right|
^{-1}   \left[ \frac{|\kappa|(2-\alpha)^2}{2m\Omega^2}\right]^{\frac{1}{2-\alpha}} \exp (\Omega \tau_s),
\end{equation}
where $\Gamma_{E}$ is the Euler Gamma function and $\Gamma_0$ is the tunneling rate in the absence of the alternating field. The time averaging of Eq. (\ref{t_dep_tunneling_enhancement_app}) gives
\begin{equation}\label{taveraged_tunneling_enhancement_app}
\overline{\Gamma(\mathcal{E})}=\Gamma_0 \frac{1}{\sqrt{2\pi a_1}} \exp(a_1).
\end{equation}
Eq. (\ref{taveraged_tunneling_enhancement_app}) shows that the semiclassical description is relevant only when
\begin{equation}
S_0\gg a_1\gg 1.
\end{equation}
If the condition $S_0 \gg a_1$ is not satisfied, in addition to linear expansion in $\mathcal{E}$, higher order corrections need to be considered.

\begin{figure}
\centering
\includegraphics[scale=0.5]{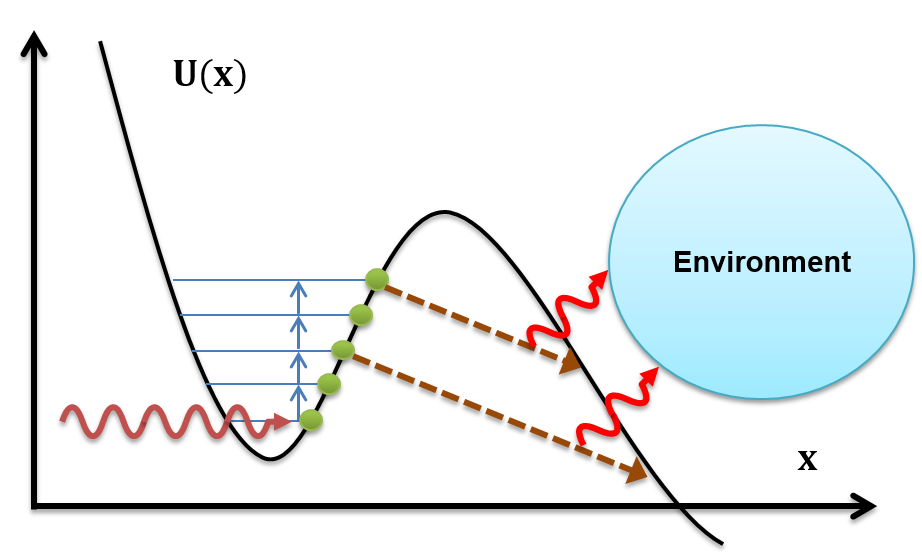}
\caption[Quantum tunneling of a particle from the metastable potential barrier under the influence of a high frequency field is shown.]{ Dissipative quantum tunneling of a particle from the metastable potential barrier under the influence of a high frequency field is shown. The incident radiation can excite the tunneling field to higher energy states with higher rate of tunneling; however, the probability for being excited to higher energy states would decreases. The dotted arrow indicates tunneling with dissipation where the field loses some energy to the environment. The emerging particle has less energy compared to the energy before tunneling and has to travel longer distance under the barrier.}\label{multiphoton_tunneling}
\end{figure}

In this semi-calssical method the probability of multi-photon processes in the enhancement of the quantum tunneling has been automatically included, \cite{PhysRevB.28.6281, SovPhysJetpIvelevMelnikov1985}. In a multi-photon process, the charge trapped in a local minima can absorb either one or more than one  photon and tunnels away. Although, absorbing more than one photons is less probable; however, upon absorption of more photons the particle becomes exited to a higher level from which tunneling is facilitated, thus  the probability of quantum tunneling increases (see Fig. \ref{multiphoton_tunneling}).

\bibliography{HighFrequencyQPSbib}

\begin{thebibliography}{42}%
\makeatletter
\providecommand \@ifxundefined [1]{%
 \@ifx{#1\undefined}
}%
\providecommand \@ifnum [1]{%
 \ifnum #1\expandafter \@firstoftwo
 \else \expandafter \@secondoftwo
 \fi
}%
\providecommand \@ifx [1]{%
 \ifx #1\expandafter \@firstoftwo
 \else \expandafter \@secondoftwo
 \fi
}%
\providecommand \natexlab [1]{#1}%
\providecommand \enquote  [1]{``#1''}%
\providecommand \bibnamefont  [1]{#1}%
\providecommand \bibfnamefont [1]{#1}%
\providecommand \citenamefont [1]{#1}%
\providecommand \href@noop [0]{\@secondoftwo}%
\providecommand \href [0]{\begingroup \@sanitize@url \@href}%
\providecommand \@href[1]{\@@startlink{#1}\@@href}%
\providecommand \@@href[1]{\endgroup#1\@@endlink}%
\providecommand \@sanitize@url [0]{\catcode `\\12\catcode `\$12\catcode
  `\&12\catcode `\#12\catcode `\^12\catcode `\_12\catcode `\%12\relax}%
\providecommand \@@startlink[1]{}%
\providecommand \@@endlink[0]{}%
\providecommand \url  [0]{\begingroup\@sanitize@url \@url }%
\providecommand \@url [1]{\endgroup\@href {#1}{\urlprefix }}%
\providecommand \urlprefix  [0]{URL }%
\providecommand \Eprint [0]{\href }%
\providecommand \doibase [0]{http://dx.doi.org/}%
\providecommand \selectlanguage [0]{\@gobble}%
\providecommand \bibinfo  [0]{\@secondoftwo}%
\providecommand \bibfield  [0]{\@secondoftwo}%
\providecommand \translation [1]{[#1]}%
\providecommand \BibitemOpen [0]{}%
\providecommand \bibitemStop [0]{}%
\providecommand \bibitemNoStop [0]{.\EOS\space}%
\providecommand \EOS [0]{\spacefactor3000\relax}%
\providecommand \BibitemShut  [1]{\csname bibitem#1\endcsname}%
\let\auto@bib@innerbib\@empty
\bibitem [{\citenamefont {Arutyunov}\ \emph {et~al.}(2008)\citenamefont
  {Arutyunov}, \citenamefont {Golubev},\ and\ \citenamefont
  {Zaikin}}]{Arutyunov20081}%
  \BibitemOpen
  \bibfield  {author} {\bibinfo {author} {\bibfnamefont {K.}~\bibnamefont
  {Arutyunov}}, \bibinfo {author} {\bibfnamefont {D.}~\bibnamefont {Golubev}},
  \ and\ \bibinfo {author} {\bibfnamefont {A.}~\bibnamefont {Zaikin}},\ }\href
  {\doibase 10.1016/j.physrep.2008.04.009} {\bibfield  {journal} {\bibinfo
  {journal} {Physics Reports}\ }\textbf {\bibinfo {volume} {464}},\ \bibinfo
  {pages} {1 } (\bibinfo {year} {2008})}\BibitemShut {NoStop}%
\bibitem [{Bez(2008)}]{Bezryadin}%
  \BibitemOpen
  \href@noop {} {\bibfield  {journal} {\bibinfo  {journal} {Journal of Physics:
  Condensed Matter}\ }\textbf {\bibinfo {volume} {20}},\ \bibinfo {pages}
  {043202} (\bibinfo {year} {2008})}\BibitemShut {NoStop}%
\bibitem [{\citenamefont {Kerman}(2013)}]{1367-2630-15-10-105017}%
  \BibitemOpen
  \bibfield  {author} {\bibinfo {author} {\bibfnamefont {A.~J.}\ \bibnamefont
  {Kerman}},\ }\href {http://stacks.iop.org/1367-2630/15/i=10/a=105017}
  {\bibfield  {journal} {\bibinfo  {journal} {New Journal of Physics}\ }\textbf
  {\bibinfo {volume} {15}},\ \bibinfo {pages} {105017} (\bibinfo {year}
  {2013})}\BibitemShut {NoStop}%
\bibitem [{\citenamefont {Mooij}\ and\ \citenamefont
  {Nazarov}(2006)}]{mooijsuperconducting2006}%
  \BibitemOpen
  \bibfield  {author} {\bibinfo {author} {\bibfnamefont {J.~E.}\ \bibnamefont
  {Mooij}}\ and\ \bibinfo {author} {\bibfnamefont {Y.~V.}\ \bibnamefont
  {Nazarov}},\ }\href {\doibase 10.1038/nphys234} {\bibfield  {journal}
  {\bibinfo  {journal} {Nature Physics}\ }\textbf {\bibinfo {volume} {2}},\
  \bibinfo {pages} {169} (\bibinfo {year} {2006})}\BibitemShut {NoStop}%
\bibitem [{\citenamefont {Mooij}\ \emph {et~al.}(2015)\citenamefont {Mooij},
  \citenamefont {Schön}, \citenamefont {Shnirman}, \citenamefont {Fuse},
  \citenamefont {Harmans}, \citenamefont {Rotzinger},\ and\ \citenamefont
  {Verbruggen}}]{MooijSchoen}%
  \BibitemOpen
  \bibfield  {author} {\bibinfo {author} {\bibfnamefont {J.~E.}\ \bibnamefont
  {Mooij}}, \bibinfo {author} {\bibfnamefont {G.}~\bibnamefont {Schön}},
  \bibinfo {author} {\bibfnamefont {A.}~\bibnamefont {Shnirman}}, \bibinfo
  {author} {\bibfnamefont {T.}~\bibnamefont {Fuse}}, \bibinfo {author}
  {\bibfnamefont {C.~J. P.~M.}\ \bibnamefont {Harmans}}, \bibinfo {author}
  {\bibfnamefont {H.}~\bibnamefont {Rotzinger}}, \ and\ \bibinfo {author}
  {\bibfnamefont {A.~H.}\ \bibnamefont {Verbruggen}},\ }\href
  {http://stacks.iop.org/1367-2630/17/i=3/a=033006} {\bibfield  {journal}
  {\bibinfo  {journal} {New Journal of Physics}\ }\textbf {\bibinfo {volume}
  {17}},\ \bibinfo {pages} {033006} (\bibinfo {year} {2015})}\BibitemShut
  {NoStop}%
\bibitem [{\citenamefont {Mooij}\ and\ \citenamefont
  {Harmans}(2005)}]{1367-2630-7-1-219}%
  \BibitemOpen
  \bibfield  {author} {\bibinfo {author} {\bibfnamefont {J.~E.}\ \bibnamefont
  {Mooij}}\ and\ \bibinfo {author} {\bibfnamefont {C.~J. P.~M.}\ \bibnamefont
  {Harmans}},\ }\href {http://stacks.iop.org/1367-2630/7/i=1/a=219} {\bibfield
  {journal} {\bibinfo  {journal} {New Journal of Physics}\ }\textbf {\bibinfo
  {volume} {7}},\ \bibinfo {pages} {219} (\bibinfo {year} {2005})}\BibitemShut
  {NoStop}%
\bibitem [{\citenamefont {Astafiev}\ \emph {et~al.}(2012)\citenamefont
  {Astafiev}, \citenamefont {Ioffe}, \citenamefont {Kafanov}, \citenamefont
  {Pashkin}, \citenamefont {Arutyunov}, \citenamefont {Shahar}, \citenamefont
  {Cohen},\ and\ \citenamefont {Tsai}}]{citeulike:10580965}%
  \BibitemOpen
  \bibfield  {author} {\bibinfo {author} {\bibfnamefont {O.~V.}\ \bibnamefont
  {Astafiev}}, \bibinfo {author} {\bibfnamefont {L.~B.}\ \bibnamefont {Ioffe}},
  \bibinfo {author} {\bibfnamefont {S.}~\bibnamefont {Kafanov}}, \bibinfo
  {author} {\bibfnamefont {Y.}~\bibnamefont {Pashkin}}, \bibinfo {author}
  {\bibnamefont {Arutyunov}}, \bibinfo {author} {\bibfnamefont
  {D.}~\bibnamefont {Shahar}}, \bibinfo {author} {\bibfnamefont
  {O.}~\bibnamefont {Cohen}}, \ and\ \bibinfo {author} {\bibfnamefont {J.~S.}\
  \bibnamefont {Tsai}},\ }\href {\doibase 10.1038/nature10930} {\bibfield
  {journal} {\bibinfo  {journal} {Nature}\ }\textbf {\bibinfo {volume} {484}},\
  \bibinfo {pages} {355} (\bibinfo {year} {2012})}\BibitemShut {NoStop}%
\bibitem [{\citenamefont {Webster}\ \emph {et~al.}(2008)\citenamefont
  {Webster}, \citenamefont {Giblin}, \citenamefont {Cox}, \citenamefont
  {Janssen},\ and\ \citenamefont {Zorin}}]{4574936}%
  \BibitemOpen
  \bibfield  {author} {\bibinfo {author} {\bibfnamefont {C.~H.}\ \bibnamefont
  {Webster}}, \bibinfo {author} {\bibfnamefont {S.}~\bibnamefont {Giblin}},
  \bibinfo {author} {\bibfnamefont {D.}~\bibnamefont {Cox}}, \bibinfo {author}
  {\bibfnamefont {T.~J. B.~M.}\ \bibnamefont {Janssen}}, \ and\ \bibinfo
  {author} {\bibfnamefont {A.}~\bibnamefont {Zorin}},\ }in\ \href@noop {}
  {\emph {\bibinfo {booktitle} {Precision Electromagnetic Measurements Digest,
  2008. CPEM 2008. Conference on}}}\ (\bibinfo {year} {2008})\ pp.\ \bibinfo
  {pages} {628--629}\BibitemShut {NoStop}%
\bibitem [{\citenamefont {Hriscu}\ and\ \citenamefont
  {Nazarov}(2011)}]{PhysRevB.83.174511}%
  \BibitemOpen
  \bibfield  {author} {\bibinfo {author} {\bibfnamefont {A.~M.}\ \bibnamefont
  {Hriscu}}\ and\ \bibinfo {author} {\bibfnamefont {Y.~V.}\ \bibnamefont
  {Nazarov}},\ }\href {\doibase 10.1103/PhysRevB.83.174511} {\bibfield
  {journal} {\bibinfo  {journal} {Phys. Rev. B}\ }\textbf {\bibinfo {volume}
  {83}},\ \bibinfo {pages} {174511} (\bibinfo {year} {2011})}\BibitemShut
  {NoStop}%
\bibitem [{\citenamefont {Hongisto}\ and\ \citenamefont
  {Zorin}(2012)}]{PhysRevLett.108.097001}%
  \BibitemOpen
  \bibfield  {author} {\bibinfo {author} {\bibfnamefont {T.~T.}\ \bibnamefont
  {Hongisto}}\ and\ \bibinfo {author} {\bibfnamefont {A.~B.}\ \bibnamefont
  {Zorin}},\ }\href {\doibase 10.1103/PhysRevLett.108.097001} {\bibfield
  {journal} {\bibinfo  {journal} {Phys. Rev. Lett.}\ }\textbf {\bibinfo
  {volume} {108}},\ \bibinfo {pages} {097001} (\bibinfo {year}
  {2012})}\BibitemShut {NoStop}%
\bibitem [{\citenamefont {Giordano}(1988)}]{gio}%
  \BibitemOpen
  \bibfield  {author} {\bibinfo {author} {\bibfnamefont {N.}~\bibnamefont
  {Giordano}},\ }\href@noop {} {\bibfield  {journal} {\bibinfo  {journal}
  {Phys. Rev. Letters}\ }\textbf {\bibinfo {volume} {61}},\ \bibinfo {pages}
  {2137} (\bibinfo {year} {1988})}\BibitemShut {NoStop}%
\bibitem [{\citenamefont {Lau}\ \emph {et~al.}(2001)\citenamefont {Lau},
  \citenamefont {Markovic}, \citenamefont {Bockrath}, \citenamefont
  {Bezryadin},\ and\ \citenamefont {Tinkham}}]{PhysRevLett.87.217003}%
  \BibitemOpen
  \bibfield  {author} {\bibinfo {author} {\bibfnamefont {C.~N.}\ \bibnamefont
  {Lau}}, \bibinfo {author} {\bibfnamefont {N.}~\bibnamefont {Markovic}},
  \bibinfo {author} {\bibfnamefont {M.}~\bibnamefont {Bockrath}}, \bibinfo
  {author} {\bibfnamefont {A.}~\bibnamefont {Bezryadin}}, \ and\ \bibinfo
  {author} {\bibfnamefont {M.}~\bibnamefont {Tinkham}},\ }\href {\doibase
  10.1103/PhysRevLett.87.217003} {\bibfield  {journal} {\bibinfo  {journal}
  {Phys. Rev. Lett.}\ }\textbf {\bibinfo {volume} {87}},\ \bibinfo {pages}
  {217003} (\bibinfo {year} {2001})}\BibitemShut {NoStop}%
\bibitem [{\citenamefont {Tinkham}(1996)}]{tinkham}%
  \BibitemOpen
  \bibfield  {author} {\bibinfo {author} {\bibfnamefont {M.}~\bibnamefont
  {Tinkham}},\ }\href@noop {} {\emph {\bibinfo {title} {Introduction to
  Superconductivity (New York: McGraw-Hill)}}}\ (\bibinfo  {publisher} {New
  York: McGraw-Hill},\ \bibinfo {year} {1996})\BibitemShut {NoStop}%
\bibitem [{\citenamefont {Ma}\ and\ \citenamefont
  {Lee}(1985)}]{PhysRevB.32.5658}%
  \BibitemOpen
  \bibfield  {author} {\bibinfo {author} {\bibfnamefont {M.}~\bibnamefont
  {Ma}}\ and\ \bibinfo {author} {\bibfnamefont {P.~A.}\ \bibnamefont {Lee}},\
  }\href {\doibase 10.1103/PhysRevB.32.5658} {\bibfield  {journal} {\bibinfo
  {journal} {Phys. Rev. B}\ }\textbf {\bibinfo {volume} {32}},\ \bibinfo
  {pages} {5658} (\bibinfo {year} {1985})}\BibitemShut {NoStop}%
\bibitem [{\citenamefont {Sadovskii}(1997)}]{Sadovskii1997225}%
  \BibitemOpen
  \bibfield  {author} {\bibinfo {author} {\bibfnamefont {M.~V.}\ \bibnamefont
  {Sadovskii}},\ }\href {\doibase
  http://dx.doi.org/10.1016/S0370-1573(96)00036-1} {\bibfield  {journal}
  {\bibinfo  {journal} {Physics Reports}\ }\textbf {\bibinfo {volume} {282}},\
  \bibinfo {pages} {225 } (\bibinfo {year} {1997})}\BibitemShut {NoStop}%
\bibitem [{\citenamefont {{Sac{\'e}p{\'e}}}\ \emph {et~al.}(2011)\citenamefont
  {{Sac{\'e}p{\'e}}}, \citenamefont {{Dubouchet}}, \citenamefont {{Chapelier}},
  \citenamefont {{Sanquer}}, \citenamefont {{Ovadia}}, \citenamefont
  {{Shahar}}, \citenamefont {{Feigel'Man}},\ and\ \citenamefont
  {{Ioffe}}}]{Sacepe2011jm}%
  \BibitemOpen
  \bibfield  {author} {\bibinfo {author} {\bibfnamefont {B.}~\bibnamefont
  {{Sac{\'e}p{\'e}}}}, \bibinfo {author} {\bibfnamefont {T.}~\bibnamefont
  {{Dubouchet}}}, \bibinfo {author} {\bibfnamefont {C.}~\bibnamefont
  {{Chapelier}}}, \bibinfo {author} {\bibfnamefont {M.}~\bibnamefont
  {{Sanquer}}}, \bibinfo {author} {\bibfnamefont {M.}~\bibnamefont {{Ovadia}}},
  \bibinfo {author} {\bibfnamefont {D.}~\bibnamefont {{Shahar}}}, \bibinfo
  {author} {\bibfnamefont {M.}~\bibnamefont {{Feigel'Man}}}, \ and\ \bibinfo
  {author} {\bibfnamefont {L.}~\bibnamefont {{Ioffe}}},\ }\href {\doibase
  10.1038/nphys1892} {\bibfield  {journal} {\bibinfo  {journal} {Nature
  Physics}\ }\textbf {\bibinfo {volume} {7}},\ \bibinfo {pages} {239} (\bibinfo
  {year} {2011})}\BibitemShut {NoStop}%
\bibitem [{\citenamefont {Caldeira}\ and\ \citenamefont
  {Leggett}(1981)}]{PhysRevLett.46.211}%
  \BibitemOpen
  \bibfield  {author} {\bibinfo {author} {\bibfnamefont {A.~O.}\ \bibnamefont
  {Caldeira}}\ and\ \bibinfo {author} {\bibfnamefont {A.~J.}\ \bibnamefont
  {Leggett}},\ }\href {\doibase 10.1103/PhysRevLett.46.211} {\bibfield
  {journal} {\bibinfo  {journal} {Phys. Rev. Lett.}\ }\textbf {\bibinfo
  {volume} {46}},\ \bibinfo {pages} {211} (\bibinfo {year} {1981})}\BibitemShut
  {NoStop}%
\bibitem [{\citenamefont {Caldeira}\ and\ \citenamefont
  {Leggett}(1983)}]{Caldeira1983374}%
  \BibitemOpen
  \bibfield  {author} {\bibinfo {author} {\bibfnamefont {A.}~\bibnamefont
  {Caldeira}}\ and\ \bibinfo {author} {\bibfnamefont {A.}~\bibnamefont
  {Leggett}},\ }\href {\doibase http://dx.doi.org/10.1016/0003-4916(83)90202-6}
  {\bibfield  {journal} {\bibinfo  {journal} {Annals of Physics}\ }\textbf
  {\bibinfo {volume} {149}},\ \bibinfo {pages} {374 } (\bibinfo {year}
  {1983})}\BibitemShut {NoStop}%
\bibitem [{\citenamefont {Bae}\ \emph {et~al.}(2012)\citenamefont {Bae},
  \citenamefont {Dinsmore}, \citenamefont {Sahu},\ and\ \citenamefont
  {Bezryadin}}]{1367-2630-14-4-043014}%
  \BibitemOpen
  \bibfield  {author} {\bibinfo {author} {\bibfnamefont {M.-H.}\ \bibnamefont
  {Bae}}, \bibinfo {author} {\bibfnamefont {R.~C.}\ \bibnamefont {Dinsmore}},
  \bibinfo {author} {\bibfnamefont {M.}~\bibnamefont {Sahu}}, \ and\ \bibinfo
  {author} {\bibfnamefont {A.}~\bibnamefont {Bezryadin}},\ }\href
  {http://stacks.iop.org/1367-2630/14/i=4/a=043014} {\bibfield  {journal}
  {\bibinfo  {journal} {New Journal of Physics}\ }\textbf {\bibinfo {volume}
  {14}},\ \bibinfo {pages} {043014} (\bibinfo {year} {2012})}\BibitemShut
  {NoStop}%
\bibitem [{\citenamefont {Lehtinen}\ \emph {et~al.}(2012)\citenamefont
  {Lehtinen}, \citenamefont {Zakharov},\ and\ \citenamefont
  {Arutyunov}}]{PhysRevLett.109.187001}%
  \BibitemOpen
  \bibfield  {author} {\bibinfo {author} {\bibfnamefont {J.~S.}\ \bibnamefont
  {Lehtinen}}, \bibinfo {author} {\bibfnamefont {K.}~\bibnamefont {Zakharov}},
  \ and\ \bibinfo {author} {\bibfnamefont {K.~Y.}\ \bibnamefont {Arutyunov}},\
  }\href {\doibase 10.1103/PhysRevLett.109.187001} {\bibfield  {journal}
  {\bibinfo  {journal} {Phys. Rev. Lett.}\ }\textbf {\bibinfo {volume} {109}},\
  \bibinfo {pages} {187001} (\bibinfo {year} {2012})}\BibitemShut {NoStop}%
\bibitem [{\citenamefont {Webster}\ \emph {et~al.}(2013)\citenamefont
  {Webster}, \citenamefont {Fenton}, \citenamefont {Hongisto}, \citenamefont
  {Giblin}, \citenamefont {Zorin},\ and\ \citenamefont
  {Warburton}}]{PhysRevB.87.144510}%
  \BibitemOpen
  \bibfield  {author} {\bibinfo {author} {\bibfnamefont {C.~H.}\ \bibnamefont
  {Webster}}, \bibinfo {author} {\bibfnamefont {J.~C.}\ \bibnamefont {Fenton}},
  \bibinfo {author} {\bibfnamefont {T.~T.}\ \bibnamefont {Hongisto}}, \bibinfo
  {author} {\bibfnamefont {S.~P.}\ \bibnamefont {Giblin}}, \bibinfo {author}
  {\bibfnamefont {A.~B.}\ \bibnamefont {Zorin}}, \ and\ \bibinfo {author}
  {\bibfnamefont {P.~A.}\ \bibnamefont {Warburton}},\ }\href {\doibase
  10.1103/PhysRevB.87.144510} {\bibfield  {journal} {\bibinfo  {journal} {Phys.
  Rev. B}\ }\textbf {\bibinfo {volume} {87}},\ \bibinfo {pages} {144510}
  (\bibinfo {year} {2013})}\BibitemShut {NoStop}%
\bibitem [{\citenamefont {Shapiro}(1963)}]{PhysRevLett.11.80}%
  \BibitemOpen
  \bibfield  {author} {\bibinfo {author} {\bibfnamefont {S.}~\bibnamefont
  {Shapiro}},\ }\href {\doibase 10.1103/PhysRevLett.11.80} {\bibfield
  {journal} {\bibinfo  {journal} {Phys. Rev. Lett.}\ }\textbf {\bibinfo
  {volume} {11}},\ \bibinfo {pages} {80} (\bibinfo {year} {1963})}\BibitemShut
  {NoStop}%
\bibitem [{\citenamefont {Anderson}(1964)}]{anderson64}%
  \BibitemOpen
  \bibfield  {author} {\bibinfo {author} {\bibfnamefont {A.~H.}\ \bibnamefont
  {Anderson}, \bibfnamefont {P.~W.;~Dayem}},\ }\href@noop {} {\bibfield
  {journal} {\bibinfo  {journal} {Phys. Rev. Lett.}\ ,\ \bibinfo {pages} {13}}
  (\bibinfo {year} {1964})}\BibitemShut {NoStop}%
\bibitem [{\citenamefont {Bae}\ \emph {et~al.}(2009)\citenamefont {Bae},
  \citenamefont {Dinsmore}, \citenamefont {Aref}, \citenamefont {Brenner},\
  and\ \citenamefont {Bezryadin}}]{Bae:2009hn}%
  \BibitemOpen
  \bibfield  {author} {\bibinfo {author} {\bibfnamefont {M.-H.}\ \bibnamefont
  {Bae}}, \bibinfo {author} {\bibfnamefont {R.~C.}\ \bibnamefont {Dinsmore}},
  \bibinfo {author} {\bibfnamefont {T.}~\bibnamefont {Aref}}, \bibinfo {author}
  {\bibfnamefont {M.}~\bibnamefont {Brenner}}, \ and\ \bibinfo {author}
  {\bibfnamefont {A.}~\bibnamefont {Bezryadin}},\ }\href@noop {} {\bibfield
  {journal} {\bibinfo  {journal} {Nano Letters}\ }\textbf {\bibinfo {volume}
  {9}},\ \bibinfo {pages} {1889} (\bibinfo {year} {2009})}\BibitemShut
  {NoStop}%
\bibitem [{\citenamefont {Ivlev}\ and\ \citenamefont
  {Mel'nikov}(1986)}]{SovPhysJETP.63.1986}%
  \BibitemOpen
  \bibfield  {author} {\bibinfo {author} {\bibfnamefont {B.~I.}\ \bibnamefont
  {Ivlev}}\ and\ \bibinfo {author} {\bibfnamefont {V.~I.}\ \bibnamefont
  {Mel'nikov}},\ }\href@noop {} {\bibfield  {journal} {\bibinfo  {journal}
  {Sov. Phys. JETP}\ }\textbf {\bibinfo {volume} {63}},\ \bibinfo {pages}
  {1295} (\bibinfo {year} {1986})}\BibitemShut {NoStop}%
\bibitem [{\citenamefont {Ivlev}\ and\ \citenamefont
  {Mel'nikov}(1985{\natexlab{a}})}]{PhysRevLett.55.1614}%
  \BibitemOpen
  \bibfield  {author} {\bibinfo {author} {\bibfnamefont {B.~I.}\ \bibnamefont
  {Ivlev}}\ and\ \bibinfo {author} {\bibfnamefont {V.~I.}\ \bibnamefont
  {Mel'nikov}},\ }\href {\doibase 10.1103/PhysRevLett.55.1614} {\bibfield
  {journal} {\bibinfo  {journal} {Phys. Rev. Lett.}\ }\textbf {\bibinfo
  {volume} {55}},\ \bibinfo {pages} {1614} (\bibinfo {year}
  {1985}{\natexlab{a}})}\BibitemShut {NoStop}%
\bibitem [{\citenamefont {Ivlev}\ and\ \citenamefont
  {Mel'nikov}(1985{\natexlab{b}})}]{SovPhysJetpIvelevMelnikov1985}%
  \BibitemOpen
  \bibfield  {author} {\bibinfo {author} {\bibfnamefont {B.~I.}\ \bibnamefont
  {Ivlev}}\ and\ \bibinfo {author} {\bibfnamefont {V.~I.}\ \bibnamefont
  {Mel'nikov}},\ }\href {\doibase 10.1103/PhysRevLett.55.1614} {\bibfield
  {journal} {\bibinfo  {journal} {Sov. Phys. JETP}\ }\textbf {\bibinfo {volume}
  {62}},\ \bibinfo {pages} {1298} (\bibinfo {year}
  {1985}{\natexlab{b}})}\BibitemShut {NoStop}%
\bibitem [{\citenamefont {Devoret}\ \emph {et~al.}(1984)\citenamefont
  {Devoret}, \citenamefont {Martinis}, \citenamefont {Esteve},\ and\
  \citenamefont {Clarke}}]{PhysRevLett.53.1260}%
  \BibitemOpen
  \bibfield  {author} {\bibinfo {author} {\bibfnamefont {M.~H.}\ \bibnamefont
  {Devoret}}, \bibinfo {author} {\bibfnamefont {J.~M.}\ \bibnamefont
  {Martinis}}, \bibinfo {author} {\bibfnamefont {D.}~\bibnamefont {Esteve}}, \
  and\ \bibinfo {author} {\bibfnamefont {J.}~\bibnamefont {Clarke}},\ }\href
  {\doibase 10.1103/PhysRevLett.53.1260} {\bibfield  {journal} {\bibinfo
  {journal} {Phys. Rev. Lett.}\ }\textbf {\bibinfo {volume} {53}},\ \bibinfo
  {pages} {1260} (\bibinfo {year} {1984})}\BibitemShut {NoStop}%
\bibitem [{\citenamefont {{Ingold}}\ and\ \citenamefont
  {{Nazarov}}(2005)}]{Ingold}%
  \BibitemOpen
  \bibfield  {author} {\bibinfo {author} {\bibfnamefont {G.-L.}\ \bibnamefont
  {{Ingold}}}\ and\ \bibinfo {author} {\bibfnamefont {Y.~V.}\ \bibnamefont
  {{Nazarov}}},\ }\href@noop {} {\bibfield  {journal} {\bibinfo  {journal}
  {eprint arXiv:cond-mat/0508728}\ } (\bibinfo {year} {2005})},\ \Eprint
  {http://arxiv.org/abs/cond-mat/0508728} {cond-mat/0508728} \BibitemShut
  {NoStop}%
\bibitem [{\citenamefont {Ansari}\ \emph {et~al.}(2013)\citenamefont {Ansari},
  \citenamefont {Wilhelm}, \citenamefont {Sinha},\ and\ \citenamefont
  {Sinha}}]{Ansari}%
  \BibitemOpen
  \bibfield  {author} {\bibinfo {author} {\bibfnamefont {M.~H.}\ \bibnamefont
  {Ansari}}, \bibinfo {author} {\bibfnamefont {F.~K.}\ \bibnamefont {Wilhelm}},
  \bibinfo {author} {\bibfnamefont {U.}~\bibnamefont {Sinha}}, \ and\ \bibinfo
  {author} {\bibfnamefont {A.}~\bibnamefont {Sinha}},\ }\href@noop {}
  {\bibfield  {journal} {\bibinfo  {journal} {Superconductor Science and
  Technology}\ }\textbf {\bibinfo {volume} {26}},\ \bibinfo {pages} {125013}
  (\bibinfo {year} {2013})}\BibitemShut {NoStop}%
\bibitem [{\citenamefont {Ansari}(2015)}]{Ansari2}%
  \BibitemOpen
  \bibfield  {author} {\bibinfo {author} {\bibfnamefont {M.~H.}\ \bibnamefont
  {Ansari}},\ }\href@noop {} {\bibfield  {journal} {\bibinfo  {journal}
  {Superconductor Science and Technology}\ }\textbf {\bibinfo {volume} {28}},\
  \bibinfo {pages} {045005} (\bibinfo {year} {2015})}\BibitemShut {NoStop}%
\bibitem [{\citenamefont {McCumber}(1968)}]{mccumber}%
  \BibitemOpen
  \bibfield  {author} {\bibinfo {author} {\bibfnamefont {D.~E.}\ \bibnamefont
  {McCumber}},\ }\href@noop {} {\bibfield  {journal} {\bibinfo  {journal}
  {Journal of Applied Physics}\ }\textbf {\bibinfo {volume} {39}},\ \bibinfo
  {pages} {3113 } (\bibinfo {year} {1968})}\BibitemShut {NoStop}%
\bibitem [{\citenamefont {Stewart}(1968)}]{stewart}%
  \BibitemOpen
  \bibfield  {author} {\bibinfo {author} {\bibfnamefont {W.~C.}\ \bibnamefont
  {Stewart}},\ }\href@noop {} {\bibfield  {journal} {\bibinfo  {journal}
  {Applied Physics Letters}\ }\textbf {\bibinfo {volume} {12}},\ \bibinfo
  {pages} {277} (\bibinfo {year} {1968})}\BibitemShut {NoStop}%
\bibitem [{\citenamefont {Larkin}\ and\ \citenamefont
  {Ovchinnikov}(1983)}]{PhysRevB.28.6281}%
  \BibitemOpen
  \bibfield  {author} {\bibinfo {author} {\bibfnamefont {A.~I.}\ \bibnamefont
  {Larkin}}\ and\ \bibinfo {author} {\bibfnamefont {Y.~N.}\ \bibnamefont
  {Ovchinnikov}},\ }\href {\doibase 10.1103/PhysRevB.28.6281} {\bibfield
  {journal} {\bibinfo  {journal} {Phys. Rev. B}\ }\textbf {\bibinfo {volume}
  {28}},\ \bibinfo {pages} {6281} (\bibinfo {year} {1983})}\BibitemShut
  {NoStop}%
\bibitem [{\citenamefont {Ivlev}\ and\ \citenamefont
  {Mel'nikov}(1992)}]{kagan1992quantum}%
  \BibitemOpen
  \bibfield  {author} {\bibinfo {author} {\bibfnamefont {B.~I.}\ \bibnamefont
  {Ivlev}}\ and\ \bibinfo {author} {\bibfnamefont {V.~I.}\ \bibnamefont
  {Mel'nikov}},\ }in\ \href {http://books.google.ca/books?id=ElDtL9qZuHUC}
  {\emph {\bibinfo {booktitle} {Quantum Tunneling in Condensed Media}}},\
  \bibinfo {editor} {edited by\ \bibinfo {editor} {\bibfnamefont
  {Y.}~\bibnamefont {Kagan}}\ and\ \bibinfo {editor} {\bibfnamefont
  {A.}~\bibnamefont {Leggett}}}\ (\bibinfo  {publisher} {Elsevier Science},\
  \bibinfo {year} {1992})\BibitemShut {NoStop}%
\bibitem [{\citenamefont {Kamenev}(2011)}]{kamenev2011field}%
  \BibitemOpen
  \bibfield  {author} {\bibinfo {author} {\bibfnamefont {A.}~\bibnamefont
  {Kamenev}},\ }\href {http://books.google.ca/books?id=mrH1mAEACAAJ} {\emph
  {\bibinfo {title} {Field Theory of Non-Equilibrium Systems}}}\ (\bibinfo
  {publisher} {Cambridge University Press},\ \bibinfo {year}
  {2011})\BibitemShut {NoStop}%
\bibitem [{\citenamefont {Zaikin}\ \emph {et~al.}(1998)\citenamefont {Zaikin},
  \citenamefont {Golubev}, \citenamefont {van Otterlo},\ and\ \citenamefont
  {Zimanyi}}]{1999cond.mat.11314Z}%
  \BibitemOpen
  \bibfield  {author} {\bibinfo {author} {\bibfnamefont {A.~D.}\ \bibnamefont
  {Zaikin}}, \bibinfo {author} {\bibfnamefont {D.~S.}\ \bibnamefont {Golubev}},
  \bibinfo {author} {\bibfnamefont {A.}~\bibnamefont {van Otterlo}}, \ and\
  \bibinfo {author} {\bibfnamefont {G.~T.}\ \bibnamefont {Zimanyi}},\ }\href
  {http://stacks.iop.org/1063-7869/41/i=2/a=A36} {\bibfield  {journal}
  {\bibinfo  {journal} {Physics-Uspekhi}\ }\textbf {\bibinfo {volume} {41}},\
  \bibinfo {pages} {226} (\bibinfo {year} {1998})}\BibitemShut {NoStop}%
\bibitem [{\citenamefont {Golubev}\ and\ \citenamefont
  {Zaikin}(2001)}]{PhysRevB.64.014504}%
  \BibitemOpen
  \bibfield  {author} {\bibinfo {author} {\bibfnamefont {D.~S.}\ \bibnamefont
  {Golubev}}\ and\ \bibinfo {author} {\bibfnamefont {A.~D.}\ \bibnamefont
  {Zaikin}},\ }\href {\doibase 10.1103/PhysRevB.64.014504} {\bibfield
  {journal} {\bibinfo  {journal} {Phys. Rev. B}\ }\textbf {\bibinfo {volume}
  {64}},\ \bibinfo {pages} {014504} (\bibinfo {year} {2001})}\BibitemShut
  {NoStop}%
\bibitem [{\citenamefont {Zaikin}\ \emph {et~al.}(1997)\citenamefont {Zaikin},
  \citenamefont {Golubev}, \citenamefont {van Otterlo},\ and\ \citenamefont
  {Zim\'anyi}}]{PhysRevLett.78.1552}%
  \BibitemOpen
  \bibfield  {author} {\bibinfo {author} {\bibfnamefont {A.~D.}\ \bibnamefont
  {Zaikin}}, \bibinfo {author} {\bibfnamefont {D.~S.}\ \bibnamefont {Golubev}},
  \bibinfo {author} {\bibfnamefont {A.}~\bibnamefont {van Otterlo}}, \ and\
  \bibinfo {author} {\bibfnamefont {G.~T.}\ \bibnamefont {Zim\'anyi}},\ }\href
  {\doibase 10.1103/PhysRevLett.78.1552} {\bibfield  {journal} {\bibinfo
  {journal} {Phys. Rev. Lett.}\ }\textbf {\bibinfo {volume} {78}},\ \bibinfo
  {pages} {1552} (\bibinfo {year} {1997})}\BibitemShut {NoStop}%
\bibitem [{\citenamefont {Jafari-Salim}(2014)}]{phdthesisamirjafarisalim}%
  \BibitemOpen
  \bibfield  {author} {\bibinfo {author} {\bibfnamefont {A.}~\bibnamefont
  {Jafari-Salim}},\ }\emph {\bibinfo {title} {Superconducting Nanostructures
  for Quantum Detection of Electromagnetic Radiation}},\ \href@noop {} {Ph.D.
  thesis},\ \bibinfo  {school} {The University of Waterloo}, \bibinfo {address}
  {Waterloo, Canada} (\bibinfo {year} {2014})\BibitemShut {NoStop}%
\bibitem [{\citenamefont {{Peltonen}}\ \emph {et~al.}(2013)\citenamefont
  {{Peltonen}}, \citenamefont {{Astafiev}}, \citenamefont {{Korneeva}},
  \citenamefont {{Voronov}}, \citenamefont {{Korneev}}, \citenamefont
  {{Charaev}}, \citenamefont {{Semenov}}, \citenamefont {{Golt'sman}},
  \citenamefont {{Ioffe}}, \citenamefont {{Klapwijk}},\ and\ \citenamefont
  {{Tsai}}}]{2013arXiv1305.6692P}%
  \BibitemOpen
  \bibfield  {author} {\bibinfo {author} {\bibfnamefont {J.~T.}\ \bibnamefont
  {{Peltonen}}}, \bibinfo {author} {\bibfnamefont {O.~V.}\ \bibnamefont
  {{Astafiev}}}, \bibinfo {author} {\bibfnamefont {Y.~P.}\ \bibnamefont
  {{Korneeva}}}, \bibinfo {author} {\bibfnamefont {B.~M.}\ \bibnamefont
  {{Voronov}}}, \bibinfo {author} {\bibfnamefont {A.~A.}\ \bibnamefont
  {{Korneev}}}, \bibinfo {author} {\bibfnamefont {I.~M.}\ \bibnamefont
  {{Charaev}}}, \bibinfo {author} {\bibfnamefont {A.~V.}\ \bibnamefont
  {{Semenov}}}, \bibinfo {author} {\bibfnamefont {G.~N.}\ \bibnamefont
  {{Golt'sman}}}, \bibinfo {author} {\bibfnamefont {L.~B.}\ \bibnamefont
  {{Ioffe}}}, \bibinfo {author} {\bibfnamefont {T.~M.}\ \bibnamefont
  {{Klapwijk}}}, \ and\ \bibinfo {author} {\bibfnamefont {J.~S.}\ \bibnamefont
  {{Tsai}}},\ }\href {\doibase 10.1103/PhysRevB.88.220506} {\bibfield
  {journal} {\bibinfo  {journal} {\prb}\ }\textbf {\bibinfo {volume} {88}},\
  \bibinfo {eid} {220506} (\bibinfo {year} {2013})}\BibitemShut {NoStop}%
\bibitem [{\citenamefont {Landau}\ and\ \citenamefont
  {Lifshit︠s︡}(1977)}]{landau1977quantum}%
  \BibitemOpen
  \bibfield  {author} {\bibinfo {author} {\bibfnamefont {L.}~\bibnamefont
  {Landau}}\ and\ \bibinfo {author} {\bibfnamefont {E.}~\bibnamefont
  {Lifshit︠s︡}},\ }\href {http://books.google.ca/books?id=J9ui6KwC4mMC}
  {\emph {\bibinfo {title} {Quantum Mechanics: Non-relativistic Theory}}},\
  Butterworth Heinemann\ (\bibinfo  {publisher} {Butterworth-Heinemann},\
  \bibinfo {year} {1977})\BibitemShut {NoStop}%
\end{thebibliography}%

\end{document}